# Choosing the Right Path for AI Integration in Engineering Companies: A Strategic Guide

Rimma Dzhusupova*[1] | Jan Bosch[2] | Helena Holmström Olsson[3]

[1]Mathematics and Computer Science, Eindhoven University of Technology, Eindhoven, The Netherlands

[2]Computer Science and Engineering, Chalmers University of Technology, Gothenburg, Sweden

[3]Computer Science and Media Technology, Malmö University Malmö, Malmö, Sweden

**Correspondence**
*Corresponding author: Rimma Dzhusupova
Email: rdzhusupova@mcdermott.com

**Abstract**

The Engineering, Procurement and Construction (EPC) businesses operating within the energy sector are recognizing the increasing importance of Artificial Intelligence (AI). Many EPC companies and their clients have realized the benefits of applying AI to their businesses in order to reduce manual work, drive productivity, and streamline future operations of engineered installations in a highly competitive industry. The current AI market offers various solutions and services to support this industry, but organizations must understand how to acquire AI technology in the most beneficial way based on their business strategy and available resources. This paper presents a framework for EPC companies in their transformation towards AI. Our work is based on examples of project execution of AI-based products development at one of the biggest EPC contractors worldwide and on insights from EPC vendor companies already integrating AI into their engineering solutions. The paper covers the entire life cycle of building AI solutions, from initial business understanding to deployment and further evolution. The framework identifies how various factors influence the choice of approach toward AI project development within large international engineering corporations. By presenting a practical guide for optimal approach selection, this paper contributes to the research in AI project management and organizational strategies for integrating AI technology into businesses. The framework might also help engineering companies choose the optimum AI approach to create business value.

**Keywords**

Machine Learning, Deep Learning, Artificial Intelligence, developing and deploying AI project, Engineering Procurement and Construction industry.

1. **INTRODUCTION**

Many scientific studies, as well as more practitioner-oriented white papers, reveal potential opportunities for adopting AI in a wide range of fields like maintenance and quality management, supply chain, engineering design (Bertolini et al., 2021), manufacturing (Fahle et al., 2020), digital marketing and healthcare (Collins et al.,2021), telecommunications (Balmer et al., 2020), weather forecasting, banking and many more (Arpteg et al., 2018). To be on the same pace with the AI industry development and benefit from emerging AI-based products or solutions, many enterprises and large corporations are reconsidering the strategic plans for integrating AI into their businesses (Siemens, 2020). However, a review of previous research in industrial AI application shows limited theoretical and empirical findings about creating business values through AI technologies (Kitsios et al., 2021). Although, there are some studies that have attempted to measure the impact of AI on business value. For example, a systematic literature review in Enholm et al., 2021 tries to explain how organizations can leverage AI technologies in their operations and reveal potential value to businesses. Other studies have explored the factors contributing to the successful adoption of AI in the industry. For example, the overview done by Hamm and Klesel, 2021 identified 36 success factors, from which 13 relate to organization, 12 to technology, and 11 to the environment. Nevertheless, there is still a limited understanding of how companies must reconsider the established workflow and business models to embrace these technologies (Pappas et al., 2018). In this work, we will particularly focus on Engineering, Procurement and Construction (EPC) industry operating in the energy sector, where the Machine Learning (ML) and Deep Learning (DL) technologies reveal high potential (Ahmad et al., 2021) but where adoption is still relatively slow compared to other industry domains (J. Lee et al., 2020).

The EPC industry involved in large-scale and complex infrastructure projects and significantly contributes to a nation's economy. However, compared to other industries, the EPC industry is relatively slow in applying machine learning and quantitative statistical analysis (J. Lee et al., 2020). Most EPC companies have various challenges such as data collection, data security or legal constraints, which are discussed in detail in (Dzhusupova et al., 2022b). Overall, this industry is considered the least digitized industry globally (Venkataraman et al., 2022). In general, AI technologies' industrial development and deployment examples are still rare and generally confined within a small cluster of large international companies (Bertolini et al., 2021). In the energy sector, especially in oil and gas, ML has only been applied to isolated tasks and has not yet gained enough trust and acceptance among those companies (Hanga and Kovalchuk, 2019). Although, EPC businesses have several strategic reasons to investigate and adopt AI. Many EPC contractors are under increasing pressure from their clients to integrate AI solutions to optimize the operation and maintenance of engineered installations. For example, many operators prefer to have AI-based solutions integrated to their plant operation system to make more accurate predictions based on manufacturer details and real-time data like operating temperatures, humidity and salt levels or other critical factors in oil and gas industry (Siemens, 2020). Based on a Siemens study that has been done in cooperation with Longitude Research Ltd. on the expansion and evolution of AI within industrial organizations (Siemens, 2020), 70% of the organizations say they do so by forming partnerships with AI specialists and consultants; 38% employ an in-house development; 31% get their AI built into vendor software; and 18% use off-the-shelf options (e.g. open-source models or AI services from cloud providers) (Siemens, 2020). However, each approach has its own pros and cons, and it is essential to realize that there might be obstacles that would influence the company's choice of one of those approaches. Therefore, it is critical for a company to select the most appropriate one for each project. Based on reports from various industries (Gartner, Stamford, 2019), (Venturebeat, 2019), (McKinsey, 2021), (Dimensional research, 2019) presenting AI applications' deployment processes and challenges, the majority of the companies are still in the pilot stage. Many companies have developed a few proof-of-concept use cases. However, they struggle to apply ML more broadly or develop complete software, including, for example, a user-friendly interface to deploy their use cases at a large scale. A recent McKinsey Global Survey found that only about 15 per cent of respondents have successfully scaled automation across multiple parts of the businesses. Furthermore, only 36 per cent of respondents said that ML algorithms had been deployed beyond the pilot stage (Panikkar et al., 2021). It is worth mentioning that there is a certain lack of clear understanding of what ML can and cannot do (Bertolini et al., 2021), which also leads to unachievable goal settings and possible failure of those projects. Meantime, the current AI market offers various partnership opportunities, ready-to-go models that only need to be re-trained for specific use cases and various services for entire AI project execution. Due to a shortage of expertise and experience, many organizations start either partnership or outsourcing of their AI projects to thousands of new AI startups. Based on the business research presented in (Panikkar et al., 2021), there are three primary options to develop the AI product as:

- build fully tailored product
- utilization of off-the-shelf solutions by taking advantage of platform-based solutions and using low- and no-code approaches
- outsourcing the work to a third party including purchasing point solutions for specific use cases.

In this paper we determine pros and cons of each approach and how they would influence the company's long-term strategy. Our contribution is a framework derived from empirical data received from four large companies operating in EPC industry of energy sector that highlights key factors like cost, schedule, business criticality or intellectual property and based on their priorities, advises the best strategy to conduct each AI project. Since the adoption of AI solutions is relatively low in EPC, particularly compared with other industries (Blanco et al.,2018), this framework can be used as a guide for an EPC companies' management in choosing the best AI project execution method considering their long-term business strategies.

This work is an extended version of a conference paper "The Goldilocks Framework: Towards Selecting the Optimal Approach to Conducting AI Projects" published at the IEEE CAIN 2022 - 1st International Conference on AI Engineering - Software Engineering for AI (Dzhusupova et al., 2022a). While the published conference paper focuses on the execution of AI projects within one company, the extended version covers interviews that provide insights into how other companies approach projects for AI product development and what factors can affect their choices. Novel results are based on the analysis of new empirical data received by interviewing EPC vendors who are already integrating AI into their engineering solutions. Based on this empirical data, we were able to develop a framework which covers the entire life cycle of building AI solutions, starting from initial business understanding to deployment and further evolution. In our work, we focus on industrial examples of developing and deploying ML/DL projects to enrich the AI experience in the EPC industry. We believe that our contributions will benefit those organizations that develop or deploy ML/DL applications for improving their digitalization capabilities and their transformation towards AI-enhanced systems development.

The paper is organized as follows: Section 2 provides an overview of AI integration into the energy sector and an introduction to the EPC industry within this sector. Section 3 highlights related work and Section 4 describes the method and data collection. The empirical findings from case companies are presented in Section 5. Section 6 concludes the research summary and presents the overall framework for optimal approach selection. Section 7 and Section 8 discusses feedback on the research questions and threats to validity, and our conclusion is presented in Section 9.

## 2. PROBLEM INDICATION AND RESEARCH QUESTIONS

### 2.1 Overview of AI integration status in the energy sector

According to the Global Industry classification standards, the Energy sector comprises companies engaged in the exploration, production, refining, marketing, storage and transportation of oil and gas, refined products, coal and consumable fuels related to the generation of energy. It also includes manufacturers that offer those companies equipment and services (MSCI, 2020), **Fig. 1** illustrates how this sector influences the economy of various countries by showing the Energy share of a country's GDP (Statista, accessed 20 February 2023a). The following **Fig. 2** provides the value of the investment in the energy sector per industry (Statista, accessed 20 February 2023b). Those statistics confirm how large the energy sector is and that it is a major recipient of investments. This is important in this research because it shows the potential impact on the global economy by AI integration in this sector in order to make it more efficient and sustainable. Currently, global energy sector shifts from fossil-based energy production and consumption systems like oil, natural gas and coal, to renewable energy sources. According to the World energy transition outlook (irena.org, accessed 22 February 2023), the most significant energy consumers and carbon emitters will have to implement their ambitious plans on making their businesses more sustainable by 2030. Therefore, governments and companies in the energy sector invest billions of dollars in technologies which help them in their energy transition phases and where artificial intelligence is starting to play a significant role (Ahmad et al., 2022). AI technology is already used to enhance the forecasting method covering applications such as product demand, personnel and inventory forecasting (Ahmad et al., 2021). In general, ML/DL-based forecasting techniques are widely used in the field of energy systems (Forootan et al., 2022). The past decade has also revealed AI's potential for inspection and maintenance applications (Koroteev and Tekic, 2021). In the field of power electronics and engineering, AI approaches such as artificial neural networks and fuzzy logic models have been widely used to optimize many technical challenges in simulation, control, estimation, and fault diagnostics (Bose, 2017). The analysis done in Ghoddusi et al., 2019 suggests that Support Vector Machine, Artificial Neural Network, and Genetic Algorithms are the most popular techniques used to predict energy prices (e.g. crude oil, natural gas, and power) and to analyze energy trends. Neural networks are used for cost estimates in the bidding proposals for EPC companies (Putra and Triyono, 2015) as well as in the supply chain where AI can support in the management of the whole supply chain process to detect potential issues and ensure timely and a high-quality project delivery (Abioye et al., 2021). In academia, studies have been done to apply Deep Learning techniques to balance power generation and consumption. In Widodo et al., 2021, authors developed a Neural Network based on LSTM architecture to predict photovoltaic generation based on the number of weather parameters. In Khan et al., 2022, authors developed a hybrid Convolutional Neural Network (CNN) and Echo State Network (ESN) model for reliable electricity generation and consumption prediction. AI also helps independent operators of large installations (e.g., offshore platforms, onshore plants, upstream and downstream oil and gas operators) to analyze and compress a large amount of data from IoT devices of complex machinery or to inspect operating installations for predictive maintenance (Koroteev and Tekic, 2021). The applications of AI within the energy sector is reflected in a Siemens survey among 515 senior leaders who are responsible for, involved in, or knowledgeable about their organization's existing or planned use of AI. **Fig. 3** (Siemens, 2020) shows their response on how and where their organizations use AI.

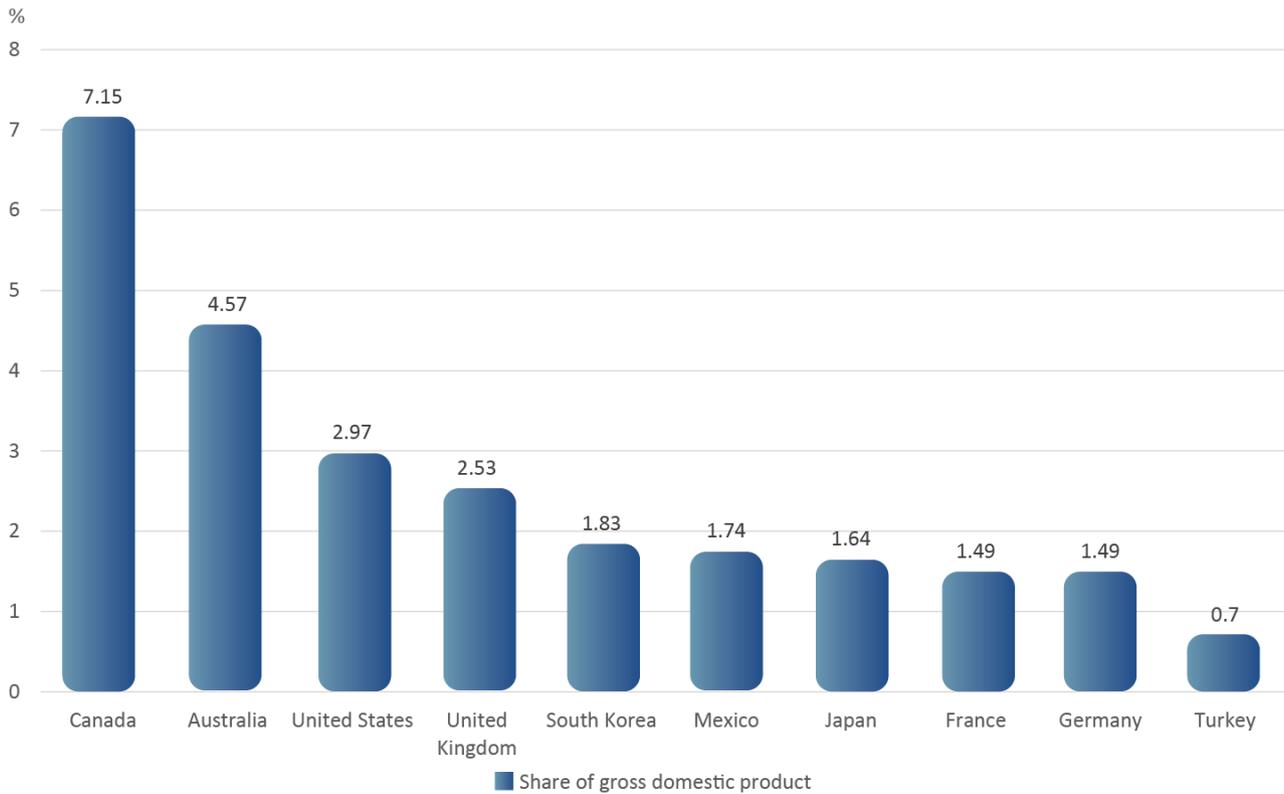

**Fig. 1 Energy-related industry value added as a percentage of GDP in 2015, by select country. Source Worldwide; OECD; 2015. Adopted from Statista, accessed 20 February 2023a.**

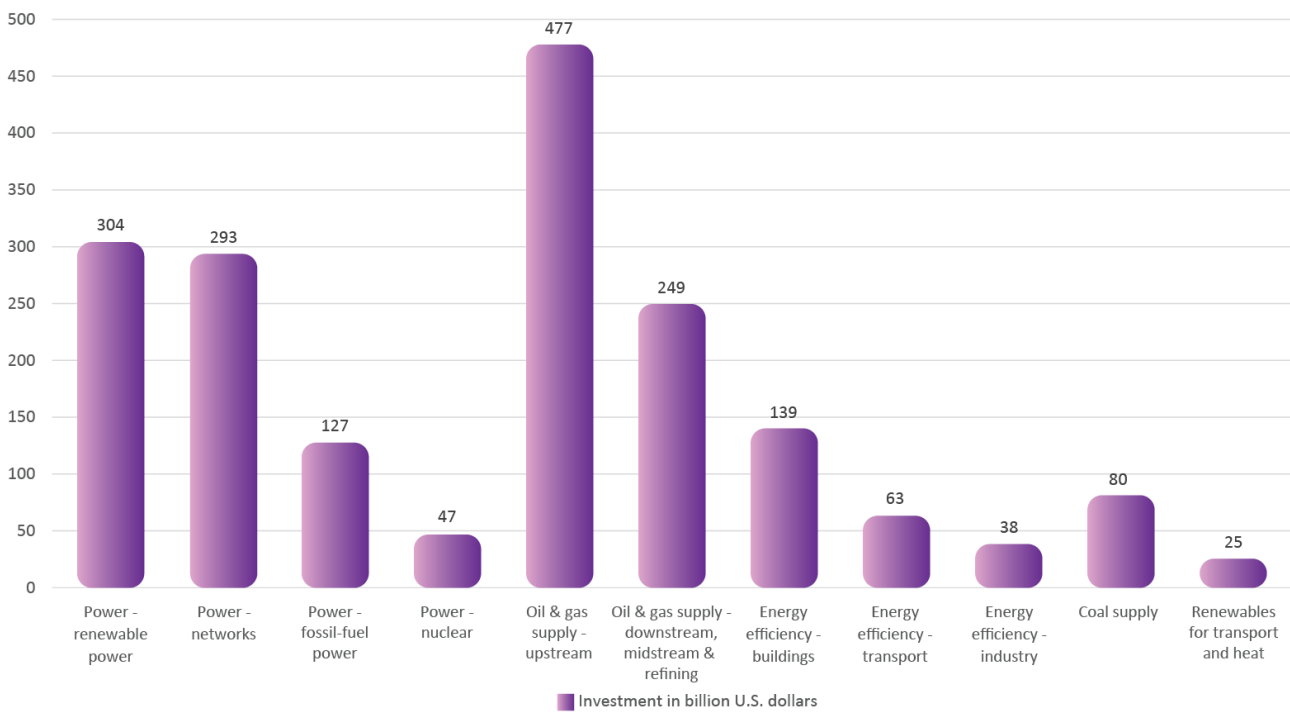

**Fig. 2 Value of energy investments worldwide in 2018, by sector (in billion US dollars). Adopted from Statista, accessed 20 February 2023b**

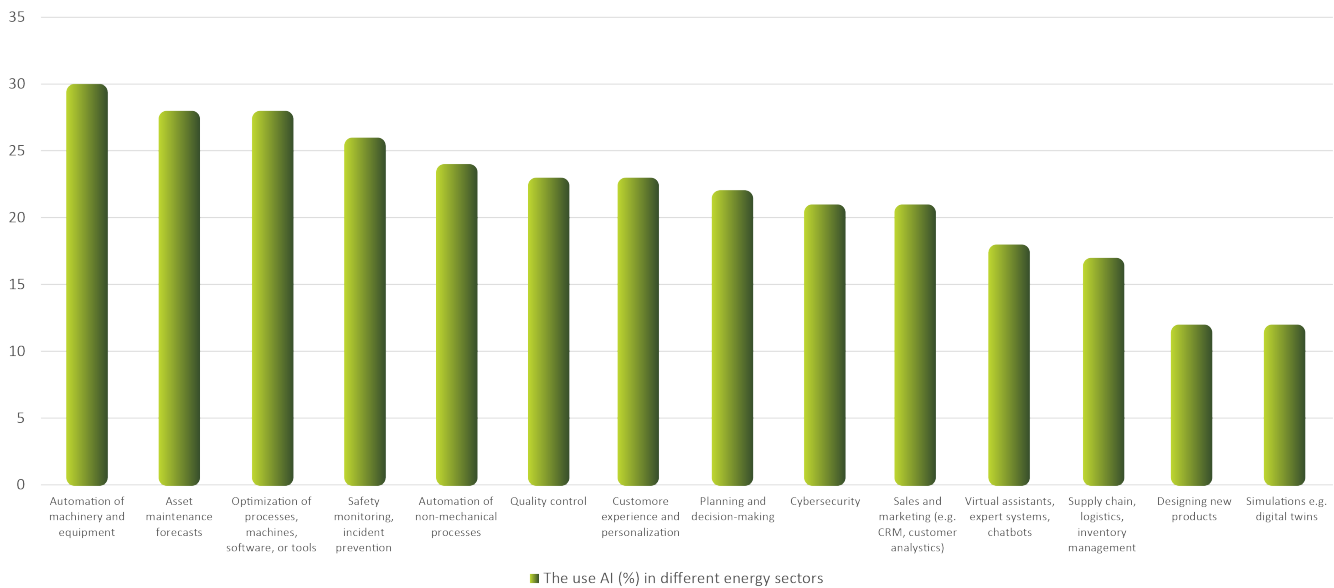

**Fig. 3 The use of AI (%) in different energy sector. Adopted from (Siemens, 2020)**

## 2.2 Introduction to EPC industry

Machine Learning and Deep Learning technologies reveal a high potential for Engineering, Procurement and Construction businesses whose projects span the entire cycle of industrial installations from bidding to engineering, construction, and start-up operation (Park et al., 2021). The typical EPC project is a technology-intensive design that relies significantly on the experience of engineering and construction teams. The rapid development of the global economy has increased the popularity of EPC projects. The need for EPC projects has been influenced by population growth, the nation's economic growth, and sustainable development concerns (Hansen, 2015). EPC projects utilize a contract-based project delivery model and are mainly applied for large-scale infrastructure works in the private sector. These are prevalent in industries such as energy or oil and gas (Sholeh and Fauziyah, 2018) where companies often rely on EPC contractors for large-scale and long-term projects that require high-skilled labor due to the sector's complexity and high safety standards (ESFC company, 2018). For example, as per IEC 61508 (an international standard published by the International Electrotechnical Commission), the minimum safety integrity level for the petrochemical industry allows only one dangerous failure in 100,000 hours (IEC, 2010).

EPC project is often associated with a "turnkey" EPC contract. The "turnkey" means that an EPC contractor holds all responsibility from the beginning of the project design until its start-up. The scope of work, in this case, includes the provision of engineering services, materials procurement and construction services (Hansen, 2015). If anything goes wrong during the project execution, a "turnkey" EPC contractor (further indicated as EPC contractor) must take care of the liabilities. Those contractors cannot surpass a guaranteed price and must complete the project delivery before the fixed deadline. If the contractors fail to meet the deadline, they are predisposed to pay the Delay Liquidated Damages (DLD) (Baron, 2018), (Blackridge research, accessed September 2022). However, the EPC project execution mode does not mean that EPC contractors must complete the entire engineering themselves. The EPC contractor can use the engineering solutions provided by the EPC vendor or subcontractor while remaining an ultimate responsible party to the client (or owner of the engineered installation) and managing vendors or subcontractors through the contract rules (Alves et al., 2016). The relationships between EPC clients, contractors and vendors are illustrated in **Fig. 4**.

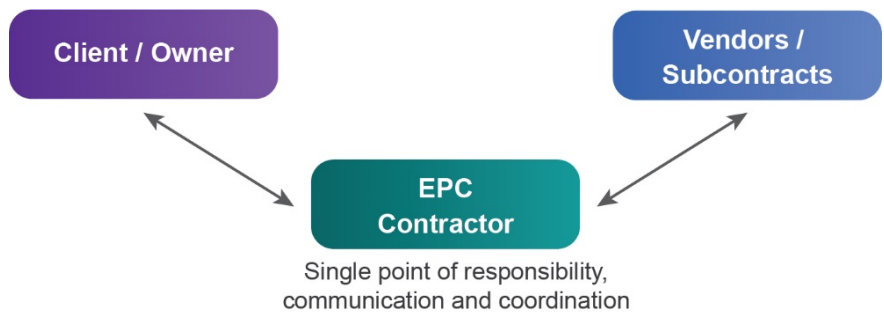

**Fig. 4 The relationships between EPC clients, contractors and vendors**

### 2.3 Research Focus

In this research, we focus on the EPC industry within the energy sector. We take as a basis the options from the business study on how companies develop AI-based solutions to generate business value (Panikkar et al., 2021). However, based on the empirical findings from this research, we extended the options of various approaches to develop the AI products. Thus, we decided to consider four approaches: (1) in-house development, (2) collaboration with a third-party specialized in AI, (3) utilization of off-the-shelf solutions including purchasing point solutions for specific use cases and (4) outsourcing the work to a third party (**Fig. 5**).

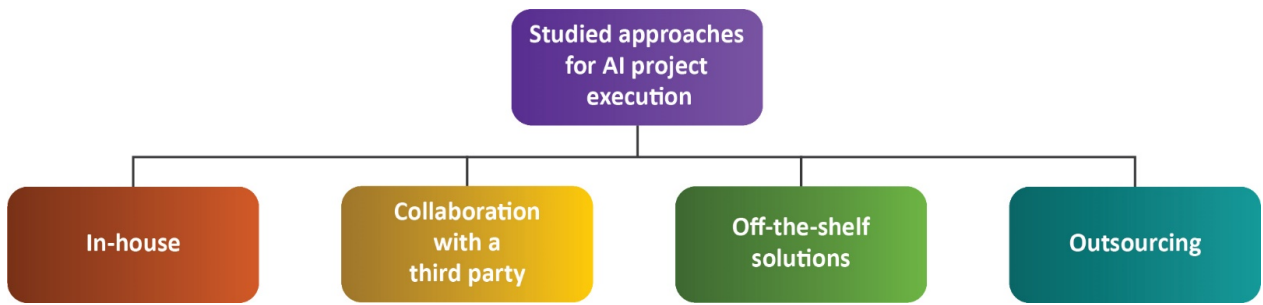

**Fig. 5 Studied approaches to develop the AI product**

Research is built on the experiences of case companies described in subsection 4.1, where case company I is an EPC company with limited experience developing and deploying AI solutions, while case companies II, III, IV and V are not turnkey contractors but specialized equipment and industrial solutions manufacturers. These companies are not responsible for the entire engineering process but provide specific components and services that are necessary for the completion of the EPC project: i.e., EPC vendors (**Fig. 4**). These EPC vendors have already successfully introduced AI in their businesses, and their experience is valuable for turnkey EPC contractors who are just starting to introduce AI in their businesses.

The research questions we aim to answer are structured in a way that differentiates between two distinct types of entities: **1.** EPC companies with limited AI experience and **2.** Companies that work with EPC companies and have more AI experience. For each group, we first identify the critical determinants and second how to approach them.

**RQ1.1: "What are the key factors that influence the development and deployment of AI solutions at an EPC company with limited AI experience?"**

This question seeks to identify the key factors influencing the development and deployment of AI solutions at an EPC company with limited AI experience. Given the relatively novel nature of AI applications in the EPC industry, there is a critical need to understand these key factors. Identifying them can provide insights into the common challenges and opportunities faced by such companies, thereby helping those companies make more informed decisions about AI adoption and implementation.

**RQ1.2: "What is the optimum approach to each key factor identified in developing and deploying AI solutions at an EPC company with limited AI experience?"**

Once the key influencing factors are identified, it becomes essential to determine the optimal approach to each. The best practices and strategies for addressing each key factor could vary greatly, depending on a variety of circumstances, such as the company's size, resources, industry positioning, etc. Understanding the optimal approach for each factor can provide a blueprint for EPC companies with limited AI experience to effectively and efficiently implement AI solutions.

**RQ2.1: "What are the key factors that influence the integration of AI solutions into businesses that work with EPC companies and have more AI experience?"**

Understanding the key factors influencing businesses that work with EPC companies and who are more experienced with AI integration of AI solutions is crucial. These companies can offer a different perspective and share lessons learned from their more advanced stage of AI integration. The insights gained can serve as valuable benchmarks for less experienced EPC companies looking to accelerate their AI adoption.

**RQ2.2: "What is the optimum approach to each key factor identified in integrating AI solutions into businesses that work with EPC companies and have more AI experience?"**

Understanding the optimal approach to each key factor identified among these more AI-experienced businesses is important. Learning how these businesses effectively navigate their key influencing factors can provide a roadmap for EPC companies. This question will help illustrate successful strategies and tactics these experienced businesses utilize, providing practical and potentially more advanced methods for integrating AI.

## 3. RELATED WORK

Although integrating AI technologies into organizational strategy is considerably more complicated with respect to other technologies (Kitsios et al., 2021), many studies have been done on analyzing different approaches of Information Technology (IT) projects' development and deployment, like outsourcing or partnering with specialized third-party companies. Similar approaches are studied in this work, and, therefore, past research on IT risks can be counted in too since the development of AI and IT projects have some common characteristics. AI development projects also require resources, requirements elicitation and development processes that are fundamentally similar to IT projects. The influence of IT on organizational performance has been widely studied (Ridwandono and Subriadi, 2019). Many frameworks, theories and methodologies have suggested the usage of IT and digital technologies, in general, in accordance with the corporate strategy (Kitsios et al., 2021) while highlighting the main opportunities and threats of outsourcing strategic IT (Verner and Abdullah, 2012), (Nie and Hammouda, 2017).

In contrary to IT, the research on industrial ML/DL application in engineering is limited. According to recent reviews in (Bertolini et al., 2021), (Paleyes et al., 2021), only a limited number of papers related to this topic were published, and the authors recognize a shortage of experience reports in the academic literature, in particular with regards to industrial development and deployment of ML/DL applications. Also, little research has been published about AI project management and organizational strategies for integrating AI technology into businesses. Many consultancy reports and roadmaps focus on different ways of AI projects deployment within companies 2020), but only a few studies advice organizations how to change their digital business strategies of incorporating AI to create new business opportunities (Mikalef et al., 2019). There are very recent publications on systematic literature review where authors focus on integration of AI into organizational strategy, highlighting the potential benefits, challenges, and opportunities (Kitsios et al., 2021), or study business value creation with AI-based technology (Borges et al., 2021). These reviews also confirm that there are still issues in practical use and a lack of knowledge regarding applying AI in a strategic way to create business value. To the best of our knowledge, there are no research papers found in academia that describe the industrial experience of AI project execution particularly within the energy sector.

## 4. RESEARCH METHOD

The method we applied in this work is based on case study research since it is appropriate for exploring real-life situations (Easterbrook et al., 2007) where unit of analysis could be a specific technology, a systems development approach or methodology, or a particular type of organization (Dubé and Paré, 2003). This research was exploratory in nature. Therefore, a case study method helped familiarize the researcher with the phenomenon itself (Saunders et al., 2016). As mentioned in Dubé and Paré, 2003, a case study approach is useful in instances when each case is unique and worth analyzing. To get an in-depth detailed exploration of the phenomena, we present multiple cases and we present each studied company as a case. In the case company I, we studied the experience of several AI projects execution within a large EPC contractor. Case companies II, III, IV and V - are interviewed EPC vendors who have already successfully introduced AI in their businesses. This method and case study selection allowed to cover the entire life cycle of building AI solutions, starting from initial business understanding until deployment and further evolution and created a basis for building a final framework. Our research utilized the inductive approach since we explored the phenomenon and built a theory in the form of a conceptual framework (Saunders et al., 2016). This approach also allowed us to study a limited number of cases with in-depth exploration (Soiferman, 2010), since the research using an inductive approach mostly focus on the context of the use case (Saunders et al., 2016). Therefore, the study of a small sample of cases while applying various techniques to collect the data to explore the phenomenon in detail was appropriate.

The summary of the approach selection at case company I, which is based on the key factors revealed during AI projects execution, are presented at the end of subsection 5.1. The key factors revealed during interviews of the case companies II, III, IV and V are highlighted in bold in subsections 5.2, 5.3, 5.4 and 5.5 and summarized in the final framework, which is presented in the form of a decision tree.

To present our empirical results, we structure AI project into a commonly used machine learning workflow across industry and research (Amershi et al., 2019), (Ashmore et al., 2019), (Hill et al., 2016), (Raji et al., 2020). This workflow is derived from prior workflows defined in the context of data science and data mining, such as TDSP (Tab et al., 2021), KDD (Fayyad et al, 1996), and CRISP-DM (Wirth and Hipp, 2000). It has the data-centred essence of the process and the multiple iteration loops among the different phases. These phases are outlined in **Fig 6**.

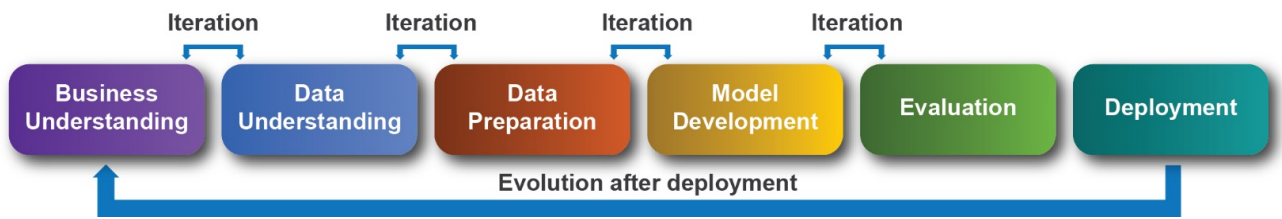

**Fig. 6: Selected workflow phases**

In the *Business Understanding* phase the business problems that AI may solve is identified, including initial decisions on what needs to be done and what resources are required to achieve it. Potential AI solutions are assessed for feasibility and added value to the company and any risks they may introduce (initial risk assessment). An initial execution plan, timeline and critical project milestones are also defined at this stage, together with the appropriate success metrics to measure how the project meets the business needs (Studer et al, 2021). In the *Data Understanding* phase, the AI models are assessed in terms of suitability and quality, and the more that is known about the data in advance, the better. Finally, it is also likely that the data owners are also project stakeholders. They may want to influence the project implementation and outcome. Therefore, they must be engaged early to avoid miscommunication and retrofits later in the project (SAIs, 2020). In the *Data Preparation* phase data is collected and prepared for AI models. This phase involves the following: exploratory data analysis, data cleaning, feature engineering, labelling, separating into training and testing sets, etc. The output of this phase is a dataset that is ready to be used in an ML/DL model. Data preparation is the first step of project execution (where the model is developed, tested and deployed). This phase aims to collect and combine all the available data and perform the necessary manipulation required as an input for the model (Ashmore et al., 2019). *Model Development and Evaluation* phases start when an AI algorithm is developed, trained and tested using the collected data. The model is then subjected to rigorous error analysis. Its performance is assessed against the appropriate metrics. *The Data Preparation and Model Development* phases are closely linked and will often have to be progressed in parallel (Studer et al, 2021). The output of this phase is an ML/DL model that has been verified and is ready to be evaluated by future product users. Developing a simple user interface is necessary to have the proof-of-concept version ready for the *Evaluation* phase to allow domain experts with no programming background to test it. Finally, evaluation of the process is required at all stages of the AI project workflow to ensure the work progresses as expected and facilitate quality control and auditing (Amershi et al., 2019). *Deployment* phase of the developed products assumes the scale-up of the product usage from a small cluster of domain experts to the entire company workforce (Amershi et al., 2019), which requires building a serving infrastructure around the model. ML code itself is a small part of real-world ML systems, but the necessary surrounding infrastructure is vast and complex (Sculley et al., 2015). The difference between a proof-of-concept model and the same model deployed at scale appeared to be very large. The models that had been developed, optimized, tested, and audited in previous phases, formed the basis of the deployment.

Once a model is deployed, it requires continuous *evolution* to maintain high-performance standards through the feedback loop (Sculley et al., 2015). During this phase, it is essential to:

- Monitor the performance of the model in operation (metrics, Key Performance Indicators - KPIs, users' feedback).
- Perform any necessary maintenance (update input data, re-train model with new data as needed).
- Identify and evaluate potential improvements based on users' feedback.

Depending on how the deployment was performed and how users' feedback is gathered and evaluated, the KPIs may or may not be the same as they were originally agreed.

### 4.1 Data collection

The data collection activities for this research lasted more than two years and are outlined in **Fig 7.** The data from case company I was being collected by the first author between 2020 and 2022. The collected empirical data is related to the experience of AI projects execution within this company and includes business case reports, execution plans and projects' close-out reports. Data from interviewed EPC vendors - case companies II, III, IV and V were gathered in the form of traditional observation by taking notes with individual observations and voice recordings. All interviews lasted between 1 and 1.5 hours. The voice recordings were transcribed for analysis by the Microsoft office transcribe feature. Below, we detail each case company, the projects we studied, and how empirical data was collected during our research.

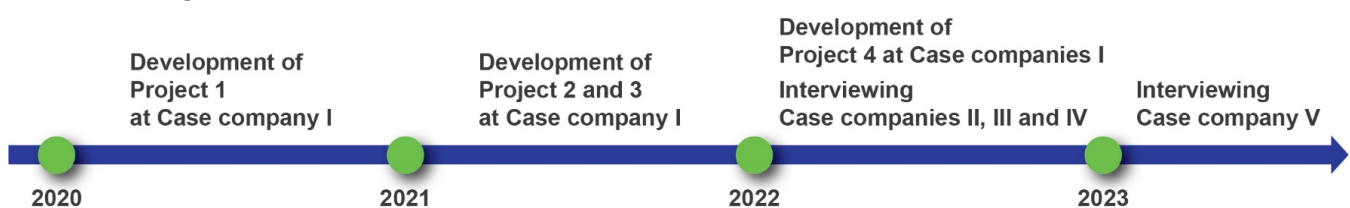

**Fig. 7 Timeline of performed activities**

**Case company I**

Case company I is a large EPC contractor operating a diversified fleet of marine construction vessels and fabrication facilities and performing oil and gas, petrochemicals and energy transition projects design and construction worldwide.

At case company I, we studied four different AI projects developed and deployed by an AI team under the supervision of the first author of this paper:

***Project 1: "Engineering hours budget prediction tool."*** It is an ML application that predicts the hours required to be spent by each engineering discipline to design a petrochemical or gas processing plant. The prediction can be made for electrical, mechanical, civil, piping, instrumentation and process disciplines. The "Engineering hours budget prediction tool" is developed in-house using data from projects executed by the company during the last fifteen years. The input data are the primary benchmark data, like mechanical equipment quantity, installed pipe length, installed power and control cables, and hardwired signals quantity. The developed Engineering hours prediction tool is used by functional management as a reference when reviewing the hours' estimate made by discipline lead engineers during the proposal phase of the project.

***Project 2: "Defect identification on the engineering drawings."*** The tool is based on the deep learning algorithm that can highlight the specific patterns on engineering drawings identified as defects or, in other words, engineering mistakes. The primary value of the Defect Identification tool is that it can assist engineers and designers with highlighting graphical design errors and, therefore, reduce manual checks. The software for operating the deep learning model has been developed by using different approaches: in-house and in collaboration with an AI company.

***Project 3: "Digitalization of PDF drawings into CAE (Computer-Aided Engineering) compatible files."*** That is an AI service provided by a third company for digitalizing engineering drawings. This service aims to support process engineers at the beginning of the project when re-drawing of the received Process & Instrumentation Diagrams (P&IDs) from the client (e.g., a petrochemical plant operator or process technology licensor) is required.

***Project 4: "Engineering knowledge database"*** The fourth use case presents an application of the solution available on the market. The provided solution can generate a comprehensive answer to a technical question based on the various engineering specifications and guidelines developed within the case company I. The model had to be re-trained with company data before evaluation and further deployment.

Table 1 illustrates which approach was chosen for each project.

| Approach | | | |
|---|---|---|---|
| **In-house development** | **Collaboration with a third-party** | **Outsourcing** | **Off-the-shelf solution** |
| Project 1 | Project 2 | Project 3 | Project 4 |
| Project 2 | | | |

**Table 1: Mapping of AI projects and utilized approaches at case company I**

The selection of case studies for case company 1 has been made based on a number of strategic factors, like data sensitivity, legal constrains, intellectual property, available resources and cost. These projects encompass a wide spectrum of AI implementation within the engineering industry, with each project offering unique insights into different aspects of AI development and deployment.

For instance, **Project 1** leverages the power of machine learning to streamline the process of resource allocation for various engineering disciplines in plant design. The uniqueness of this project stems from its application in reducing inefficiencies in the engineering budgeting process and where the data sensitivity used for model training affects on the entire project execution approach.

**Project 2** uses deep learning algorithms for quality control in engineering designs, showcasing how AI can enhance the accuracy of error detection in engineering drawings. It is distinct from software development projects, as the tool is focused on identifying graphical errors, which necessitates unique algorithms for image processing.

However, the utilization of historical project data for model training raises concerns over data ownership and contract violations. This issue underlines the importance of involving the legal department early on to ensure contractual compliance when third parties are engaged in product development.

**Project 3** showcases how AI can be employed in digital transformation initiatives within engineering businesses. Here, AI is not just utilized as a tool for automation but for transforming traditional workflows, making it distinct from generic data engineering projects. Again, the sensitive nature of the drawings' content demands cautious handling of contractual obligations and non-disclosure agreements.

Finally, **Project 4** showcases the application of AI in knowledge management within the engineering domain. It highlights the potential of AI in extracting and synthesizing technical information from diverse sources, offering value beyond conventional software development projects.

The EPC industry is still relatively inexperienced in the application of AI technologies, and there is a scarcity of readily available off-the-shelf solutions that can seamlessly fit their needs. Consequently, tailor-made solutions become a necessity. However, for an EPC company, software development is not a core business activity, and they may not be fully equipped for the development, operation, and maintenance of their AI-based products. Similarly, the issue of legal constraints surfaces. Existing contracts may not be up-to-date and comprehensive enough to cover potential instances of sensitive data leakage if third-party AI solutions are integrated into their engineering work. This situation presents a "grey zone" of potential legal risks. With the industry lacking substantial experience and precedents concerning contractual violations due to the introduction of AI, uncertainty increases. These case studies elucidate these implications, demonstrating how companies have attempted to navigate these complexities by opting for varied approaches in their AI development and deployment strategies.

**Case companies II, III, IV and V**

To extend our research to more companies we chose to interview EPC vendors since, contrary to EPC contractors, they have the experience of the entire life cycle of the AI project development. Interviewed EPC vendors are vendors of the case company I and follow the same requirements for safety for the engineered products and solutions (IEC, 2010). These vendors provide products or entire solutions which case company I is integrating into the final design of the engineered installation, such as a petrochemical plant.

Case company II is a multinational company specializing in digital automation and energy management. It combines technologies, real-time automation, software, and services. It specializes in commercial and non-commercial buildings, data centres, and various industries, including the energy sector. It is also a research company.

Case company III is the one of largest pump manufacturers in the world. The company produces pump units, electric motors, and complete solutions to control pumps and other related systems.

Case company IV is a multinational company that specializes within the energy sector in energy management, smart grid and power distribution equipment. It combines technologies, industrial automation and control, software, and services. It is also a research company.

Case company V is an industrial software developer. Its solutions address complex environments where it is critical to optimize the asset design, operation and maintenance lifecycle.

The interview protocol was designed to gather in-depth insights into the management of projects associated with AI product development, focusing on our research questions. The selection of participants was targeted at professionals who play a decision-making role in the AI integration strategy within their respective organizations. The roles of the interviewed people are presented in Table 2.

| Role within organization | Case company |
|---|---|
| **SVP, Chief Artificial Intelligence Officer** | Company II |
| **VP, AI Solutions** | Company II |
| **Lead Data Scientist. AI Solutions** | Company III |
| **Head of Research Group** | Company IV |
| **Lead Research Scientist** | Company IV |
| **SVP, Artificial Intelligence Technology** | Company V |

**Table 2: Participants roles at case companies**

The decision to select only one or two participants from companies was dictated by the roles they occupied - all individuals are AI leaders in their respective organizations. Thus, their insights were considered as highly valuable and representative of the AI-related decision-making processes in their companies.

Interview questions were framed to elicit information regarding the factors influencing the choice of different AI approaches, the potential long-term impact on business strategy, and the organization's general philosophy towards AI adoption. The interviews were structured as per ML workflow phases illustrated in Fig 6. This protocol allowed authors to comprehensively understand the varying strategies across different organizations.

Each interview lasted approximately 60 to 90 minutes in the style of informal discussions, taking notes casually but having the whole conversation recorded. Transcripts from the voice recording have been later generated for further analysis using Transcribe feature in Microsoft Word.

### 4.2 Data Analysis

The data gathered during the interviews has been analyzed using the thematic analysis technique. This method involves identifying and categorizing recurrent patterns in the data (Braun and Clarke, 2006) and can also be applied to the data gathered through interviews by transcribing the recording of interviews, reading through the transcripts multiple times, and coding the data into meaningful categories (Kiger and Varpio, 2020). The analysis was conducted by the first author, following a rigorous, systematic approach comprising six steps outlined by Braun and Clarke, 2006:

- *Step 1 - Familiarizing with the data.* This initial step involved carefully reading through each report and transcript of voice recordings, ensuring a robust understanding of the content.

- *Step 2 - Generating initial codes.* The next phase was to assign labels to words, phrases, sentences, or sections that were deemed relevant, utilizing color coding for organization and easy reference.

- *Step 3 - Searching for themes.* The data was then conceptualized, creating categories and subcategories by grouping the codes established in the preceding step.

- *Step 4 - Reviewing the themes.* This step involved connecting the categories and subcategories generated earlier, looking for coherent patterns and relations between different themes.

- *Step 5 - Defining and naming themes.* Further analysis was conducted to determine whether a hierarchy existed among the categories, allowing for the final definition and naming of the themes.

- *Step 6 - Producing the report/manuscript.* The final step was to synthesize the findings into a framework, which was then summarized in the report.

Any conflicts that emerged during the coding process were discussed among the authors until a consensus was reached. Theoretical saturation was deemed to have been reached when no new codes or themes were emerging from the data, indicating that our data collection was sufficient for the purpose of our study.

# 5. EMPIRICAL FINDINGS

## 5.1 Case Company I

This section explores and presents four AI projects developed by the case company I. To justify the selection of the approach (**Fig. 5**), case company I identified the key criteria and the approach for each criterion which was aligned with company's business strategy in the long term.

We structured the development process into a set of project execution phases (**Fig. 6**) to underline the specifics of each project and its findings per phase.

***Business Understanding phase***. Based on the business problem and available resources, case company I performed an initial cost-benefit to make a "Go/No-Go" decision during *Business Understanding* phase. The details of cost benefit analysis are presented in Table 3.

| Project 1 | The most significant amount of work was related to data collection and processing since actual data of executed engineering projects needed to be gathered from all international offices of the company. The data gathering activity could be performed only internally. Performing this project in-house was a straightforward decision since required resources were available with the company. Data collection and processing tasks were discussed during this phase to ensure that the company had the necessary resources to complete the project. For the cost-benefit analysis, the cost of engineering hours that could be potentially saved after deployment of the application was compared with the engineering hours needed to develop this product. |
|---|---|
| Project 2 | The critical part of the work was to identify typical mistakes on the engineering drawings that require a manual check and prepare a large enough data set for supervised learning. Since the graphical symbology on engineering drawing can vary from project to project, it was essential to involve domain expertise. The data used for the dataset were sensitive and contained a lot of the company's client's know-how of technological processes. Therefore, the company decided to collect data and create the dataset in-house. However, since it was a first-of-a-kind project, it was unclear whether the company could have enough resources to build an overall software to deploy the deep learning model. Therefore, case company I decided to collaborate with a third party to build an infrastructure to host and operate the model. For the cost-benefit analysis, the cost of engineering hours that could be potentially saved after deployment the tool was compared with the investment in development. |
| Project 3 | This project utilized an AI service available on the market to outsource the work to improve the overall schedule of the EPC project execution. The cost-benefit analysis was based on a service costs vs hours that engineers would spend to complete this activity manually. |
| Project 4 | This project utilized an off-the-shelf solution with ready-to-go models that need to be re-trained using company data. The cost-benefit analysis was based on the evaluation of the pilot project. A limited number of documents has been provided to the application developers to check the initial performance of the tool and whether it can help to reduce the time engineers need to spend in order to find the correct answer through company documentation. |

**Table 3: Business Understanding phase's specifics for each AI project**

The rationale for using the cost of engineering hours that could be potentially saved after deployment of the products described in Table 3 is that it is the most direct measure of the business value acknowledged by the case company I management to estimate the return on investment. By measuring the cost of engineering hours that could be saved, a clear indication of the potential cost savings by deploying these applications could be achieved. It was considered as more direct value measure than, for example, prediction accuracy or speed, which are more abstract and might not be as indicative of the actual business value of the tool.

***The Data Understanding phase*** has been performed after project received an approval to proceed based on the cost-benefit analysis described above. The details of required data per each studied AI project are presented in Table 4.

| **Project 1** | The data owner played a crucial part since the data was very sensitive, only a handful of people were granted access. |
|---|---|
| **Project 2** | This project was based on specific patterns recognition on the drawings that were classified as engineering mistakes. Since the company aimed to develop an application that could be applied to any engineering project, it was essential to consider all possible variations of the symbology of the items – components of the pattern. Within the case company I, the symbology representation was heavily dependent on who the project's client was and whether a client had its own rules on how to represent each item on the drawing. |
| **Project 3** | Since the AI service for digitalization of the engineering drawing was intended to provide the service to each engineering project individually for a fixed cost, there was a fixed symbology for each item representation captured in the project legend table. The set of drawings required to be digitalized together with the symbology table was given to the AI service provider for their further handling. |
| **Project 4** | The input data for this project was the large set (more than 1000 text files) of specifications and guidelines developed within case company I by capturing the best engineering practices and lessons learned which had been validated by dedicated subject matter experts. All these documents were under case company ownership and stored in the structured format in the engineering library database for internal use. |

**Table 4: Data Understanding phase's specifics for each AI project**

The specifics of ***Data Preparation and Modelling phases*** are presented in Table 5 and Table 6.

| **Project 1** | In this project the definitions of provided data items were not always consistent since the data were gathered from various international offices. Efforts were made to investigate the data quality and improve it where possible. Since data was gathered in tabulated data format, a regression model was good enough to make a prediction with a required quality. |
|---|---|
| **Project 2** | Since the project team had to prepare the dataset by labelling several thousand drawings, they involved many drafters from different offices, which led to poor quality of the labelling. Each item on the drawing has a tag number or numeric value, indicating specific parameters like the angle of the valve position. Sometimes, those values were too close to the symbol and required very accurate labelling to eliminate numeric values or tag numbers within the label boundaries. At first, there was a lack of consistency and accuracy in the labelled dataset, which led to the repetition of the whole exercise. Moreover, since the engineering drawing contained the company or its client's engineering know-how, it was not allowed to use third-party labelling software to avoid data leakage. Creating labelling software was not predicted and negatively affected the project's schedule. |
| **Project 3** | In this project, the AI service provider performed a detailed data preparation phase without the company's involvement. Case company I discussed and agreed only on the output quality and timing. |
| **Project 4** | The engineering documentation was easily available in a structured format. However, the model developers had to transfer text documents into a format acceptable for training the model. |

**Table 5: Data Preparation phase's specifics for each AI project**

| **Project 1** | Modelling was a very smooth activity since a linear regression model type was used. Therefore, building the user interface was a straightforward exercise. |
|---|---|
| **Project 2** | This project faced various technical challenges during the DL model and user interface development. The DL model was heavy, and it took a long time to process the input documentation. Additional support from software front-end developers was required to build a simple user interface that had call-in functions to the DL model. |
| **Project 3** | This phase was not applicable for a project 3 |
| **Project 4** | This phase was not applicable for a project 4 |

**Table 6: Modelling phase's specifics for each AI project**

*Evaluation phase*. The developed models' performances evaluation were done by a very limited group of people identified as future users of the developed product. The evaluation phase has been performed in-house and their specifics are presented in Table 7.

| | |
|---|---|
| **Project 1** | The output of this project, which predicted the engineering hours budget for projects, was evaluated on projects' data which were not part of the training or testing dataset. |
| **Project 2** | The performance of developed model, which identified engineering mistakes on the drawing, was tested by a group of process engineers on a new set of engineering drawings. Its performance was compared with a manual quality check by domain experts. |
| **Project 3** | In project 3, quality control was part of the scope of the AI service provider. As for the service evaluation, the time required for the AI service provider to digitalize the drawings was compared with the time required for manual re-drawing of the given engineering diagrams. The main objection was that AI service costs should be less than the money value of saved engineering time. |
| **Project 4** | The application was supposed to provide a comprehensive answer to the question based on documents the model had been trained on. Domain experts reviewed the answers to see whether there were indeed answering the question and whether the information was true. |

**Table 7: Evaluation phase's specifics for each AI project**

*Deployment.* A great deal of the work laid in a software development to meet the requirements of data owners related to accessibility, security and graphical user interface. Before launching the developed products on a company scale with many multinational branches, software development was unavoidable. For **project 1**, it was possible to create the software in-house with minimum resources since the intelligence was static and built into the product. However, the product developed by **project 2** was built with a server-centric intelligence which needed to be periodically updated. In this case, the cost of the servers, intelligence runtime, and intuitive user interface for product users and model creators were essential components which were not foreseen during the initial project planning. It became an important lesson learned for the future.

*Evolution.* Due to the nature of the data updates, the evolution of researched projects are lengthy processes. That is because the company focuses on large-scale EPC projects which schedules have a minimum duration of 2-3 years and include the engineering, unique equipment manufacturing and construction phases. This phase has not been achieved during this research time frame.

### 5.1.1. Summary of the key factors and the optimal approach for each of them at case company I

As a first step, case company I identified the key questions which they faced during projects development described in **Table 3** through **Table 7** and found essential from company's AI strategy perspective. As a second step, the project teams at the case company I tried to predict the long-term consequences of various approaches to each question. These questions are summarized as key factors. These key factors and identified preferred approach for each of them are listed below:

**Intellectual property.** The in-house development approach would be the best since it ensures full rights on IP. In the case of a collaboration with a third-party, it is also possible to keep IP rights. However, it will be heavily dependent on the business strategy of the collaborator company.

**Cost-time-resources.** Outsourcing does not require any resources from the client company and could be the best solution for a short term or when the level of urgency is high. The AI service provider and the client company need to agree on the time required for the indicated scope of work and the cost.

**Performance.** It is challenging to predict the performance of AI products. If the AI project is the first of its kind for the company, then in-house development has the least predictable performance. In the case of outsourcing, AI service providers can include quality control into the overall cost. The quality control might be automatic or manual by AI service providers. Therefore, this approach would be the best if the performance is the priority.

**Legal constraints and data sharing avoidance.** There is always a risk of disclosing the information, which is the company clients' intellectual property. The existing contracts might not cover the possible sensitive data leakage if an engineering contractor involves third parties to apply AI solutions for their work. Thus, some danger of stepping into a "grey zone" exists. Therefore, the best approach to avoid any data leakage and contractual dispute in future is to use an in-house approach.

**Tailored made solutions.** A lot of domain and company-specific knowledge is required to develop a successful product which is unique for a particular company. Based on the projects 1 and 2 results, the in-house and collaborative approaches work well to meet this goal.

**Company branding.** Since various industries have started to include AI in their road maps (Balakrishnan et al., 2020), the fact that an engineering company invests in developing and deploying AI products might create a good advertisement for this company in front of its clients. At case company I, the partnership with an AI company also re-enforced the company's branding among other AI practitioners and academia.

**Internal know-how development.** Raising own subject matter experts who know in-deep the company specifics and at the same time have hands-on experience in AI development is very crucial to develop sustainable AI expertise in the long term. Investment in own resources is the best way to achieve it.

**Operation & maintenance.** The best way is always to do the operation and maintenance of the model using company resources, regardless of whether infrastructure around ML/DL model is built in-house or in collaboration with a third party (e.g., re-training with new data or re-validating the model performance). Off-the-shelf solutions might assume the constant dependence on the license, although the license cost might be neglectable compared to benefits received. In the case of outsourcing, there will be a continuous dependency on the service provider.

**Scale-up opportunity.** Off-the-shelf solution usually has a complete infrastructure built around the model and can ensure the user interface users and AI experts who will re-train and evolve the model. Based on the projects development experience, an off-the-shelf solution was the best approach for case company I to further scale-up the AI project.

The **Fig. 8** summarizes the key factors and the selection of the approach by mapping each key factor to the preferred approach.

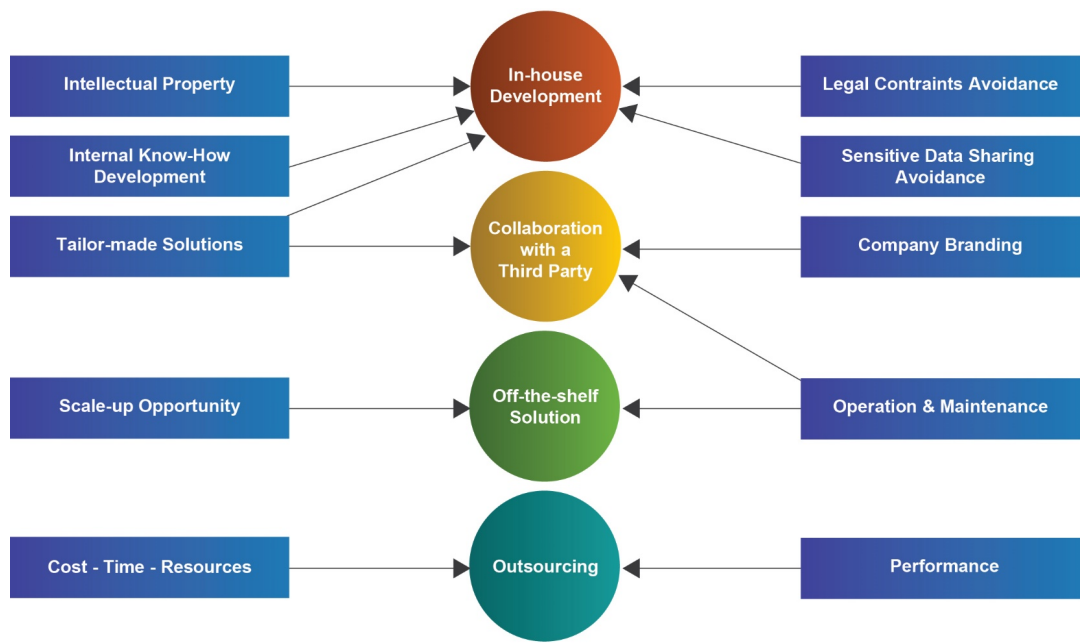

Fig. 8: Summary of the approach selection at case company I

### 5.2 Case Company II

The empirical findings from the following case are based on interviews with the Chief Artificial Intelligence Officer and the VP for AI Solutions from a multinational company specializing in digital automation and energy management.

**Business Understanding phase**

During the Business Understanding phase case company II identifies opportunities based on the company strategy. An essential requirement is that the **future product shall be unique**. Whether the product is business-critical or not, the approach in the business phase may vary. In most cases, the decision-making process is done by an in-house squad team of experts in AI and business development. If the discussed future product is not a market breaker and, therefore, not confidential, the team might also decide whether the customers could be included in the decision-making process. For significant investment, the company also utilizes external consultancy services like Accenture or BCG to ensure the external view to estimate the market demand and investment cost.

According to the VP for AI Solutions, several main criteria must be considered during the Business Understanding phase that influence which approach to choose for product development. Firstly, if there is a **general maturity of the industry** for AI applications, like predictive maintenance, for example. In this case, the product does not need to be rebuilt from scratch but only to be modified with domain specifics. In such cases, a partnership with a third party or using off-the-shelf solutions might be the best way to approach AI product development. Secondly, whether the product is **highly specialized**, so only a handful of professionals can develop it. Another criterion is what is the **time to market**, i.e. the urgency in terms of the time to market. If the urgency is high, it is easier to partner with someone who has already done it rather than build it from scratch. However, in this case, the question of **intellectual property** is rising. If the IP is valuable and the company can capitalize on it in future, then a strong preference is to keep the development in-house.

> *"Looking from multiple angles on how to engage is almost like building a Decision tree"* –
> *Vice President, AI Solutions*

In summary, the key criteria for the Business Understanding phase highlighted during interviews at case company II are:

- *Is the product for an internal or external client?*
- *What is an intellectual value (meaning business value)?*
- *Can the company technically do it?*
- *Can this product serve to company's customers?*
- *What is the timeframe for creating the product?*
- *Is there a unique competitive advantage?*
- *Is the product strategic to the company's business?*
- *Is there a strong market potential?*
- *What is the maturity of the technology?*
- *Is the future product highly specialized?*
- *What is the urgency in terms of the time to market?*

**Data Understanding and Preparation phases**

To understand what data is required for an AI project, it is crucial to have deep knowledge about the data, the customer and customer's expectations – gaining this deep knowledge is only possible by working at the company and with its customers. However, data preparation activities in case company II, in most cases, are outsourced. For example, outsourcing manual work, like labelling, to a third company with a lower hourly rate is much cheaper than doing it in-house. Case company II has a large number of AI projects running in parallel. Therefore, the outsourcing procedure is heavily standardized. The company has a list of preferred AI service providers with whom they have already signed contracts covering IP protection, data security, non-disclosure agreements, etc.

> *"If you want to be at scale, you cannot do these things project to project. We would spend too much time getting every single contract signed. We selected number of companies who are qualified to do required services and have all agreements with them in place"* – *Senior Vice President, Chief Artificial Intelligence Officer.*

**Model Development phase**

*In-house development.* Case company II has experience in using all four approaches for developing the model. However, the preferred one is to do it in-house. One of the main reasons is that company is very keen on having all the **know-how** and **IP** with themselves. They believe that by doing so, the company have a much stronger position in the market. Since many projects are running in parallel, it might happen that in case of lack of in-house developers, company hires temporary contractors for a specific scope of work. Nevertheless, this approach is still treated as an in-house development. Another reason for internal product development is that, contrary to traditional software, suitable AI-based solutions do not yet exist in the market since the business is very specific. Although, the company might consider buying a solution if it sees that the market offers a good product. However, they try to limit it. According to interviewed senior vice president, 95% of the products for external clients are developed in-house.

> *"We cannot be a differentiator on the market if we always buy instead of developing solutions internally"* – Senior Vice President, Chief Artificial Intelligence Officer.

*Partnering with a third party.* Case company II considers partnerships with AI start-ups for a **highly specialized application** where partners have unique experience and knowledge to accelerate product development. However, the partnership with EPC contractors that integrate the products in the overall design of large installations is also important.

In cases when an AI project is based on **innovative technology** which requires research, it is a common practice to partner with academia. The general preference of case company II is to create and protect IP, and for academia is to publish the findings. Therefore, the way of protecting IP is not standardized and to be considered in such partnerships case by case.

*Off-the-shelf AI solutions*. In contrary to software which could be plug-and-play solutions, AI-based solutions need a conceptualization. Therefore, utilizing ready-to-go models and re-training them is always considered as an option in case company II, especially in computer vision or neuro-linguistic programming fields. In this case, company considers utilizing off-the-shelf algorithms provided by IT giants like Google, IBM, Amazon or various AI start-ups. Although, this approach **does not apply to a critical part of company's product technology**.

> *"If you have something that is readily available and works in a very specific context that we are interested in, we can have the partnership to go to market with this off-the-shelf product provider, but then we also have to assess whether this is a long-term strategic asset that we need"* – Vice president, AI solutions.

*Outsourcing* is a common practice for **internal clients** within a company when AI solutions are aimed to help specific workflow. Internal clients, in these cases, could be HR or marketing departments. The usual strategy is to use external service providers who can deliver a specialized solution for required areas that are not a core business of case company II. Data security and privacy are crucial questions that must be considered in these cases. However, as per the interviewer's opinion, many specialized businesses can provide solutions without jeopardizing those important aspects.

**Evaluation phase**

At case company II, an internal evaluation process takes place after each product development phase. The evaluation after each phase gets more precise, and after the product has been created, the project team, consisting of AI and business development, decides whether the product is good to go to the customer. Same team might decide whether the potential customer needs to be involved in the evaluation and at which point. In developing a new AI product, a schedule is a priority at case company II. The Minimum Viable Product (MVP) needs to be created fast to reflect the urgency of the market instead of waiting for several years until the product is perfect. Case company II prefers to bring MVP as soon as it is ready to the market and see whether customers would buy it. The feedback from the market is an essential evaluation criterion to continue or not with further product improvement.

> *"So, instead of doing something perfect in two years, we do a minimum viable product with nothing more than what is needed to, then we go to a first customer and sell it"*
> *– Senior Vice President, Chief Artificial Intelligence Officer.*

**Deployment**

According to the interviewees' feedback in case company II, an off-the-shelf solution like cloud services is the best way to deploy AI products since building IT infrastructure and procuring computation capacity is not the company's core business. Current market offers complete solutions to support AI project development and deployment, which makes it unnecessary to build IT infrastructure to deploy, operate and maintain AI models at scale.

### 5.3 Case Company III

The empirical findings from the following case are based on interview with Lead Data Scientist for AI Solutions from one of the large pump manufacturers who provides equipment and entire solutions for equipment control.

**Business Understanding phase**

During the Business Understanding phase, case company III evaluates the idea with their client's involvement. They organize interview sessions with potential end users to hear their feedback. After that, the decision-making team and AI experts analyze whether AI-based solution can help realize the idea and whether it is aligned with the **company's business strategy.** They also consider the **maturity of the technology** and perform market screening to check whether the **potential solution is unique**, and no similar solutions are already available on the market. Their practice is to focus on a minimum viable product that can be built within 4-5 months and shared with end users for evaluation. To proceed with MVP, the decision-making team identifies the potential customer to engage with him and get their data for product development. The simple and lean implementation for MVP is enough to get user feedback. All the resources required for MVP are being estimated during this phase too. Case company III's usual approach is to invest its own resources and engage the client only for data collection and evaluation. There is a standard checklist which company follows during this phase, and which covers the questions as:

- *Is the problem clear?*
- *Does the company know what to do to create a solution?*
- *What is the level of maturity of the technology?*
- *Who are the stakeholders?*
- *Who is the data owner (ex., end user or third party)?*
- *What regulations need to be followed?*
- *Are there any security considerations?*

**Data Understanding and Preparation phases**

Case company III usually works with a potential customer to get the data for analyzing. For example, they connect to a customer's control system to get all the required data points, and then try to develop machine learning-based solutions with interactive visualization. If the future product is integrated into the customer's ecosystem, complying with cyber security requirements becomes a major data collection criterion.

**Model Development phase**

<u>*In-house approach.*</u> Case company III starts a development from Proof of Concept (PoC), from which they extract the requirements for future AI solutions. The PoC is usually done entirely in-house. Next, the PoC model is further developed into a final product to be ready for integration into the digital solutions which case company III offers to its clients. Thus, AI product becomes an additional feature of the software solutions company offers. In this phase, the AI team uses only their own resources and interfaces with the software department responsible for a final solution that AI-based features would be part of.

*Partnering/ Off-the shelf solutions.* If the company sees the need to implement a specific AI-based solution that already exists on the market, they could consider a partnership, especially if the solution's IP is protected. They can also consider acquiring the IP owner if it is aligned with the business strategy. According to the lead data scientist, it is a common practice in the company.

> *"If another company has developed the required AI solution, then we would go and talk to them and then just buy it instead of developing it. We do also partner with them or acquire these companies. We actually acquire start-up companies each year". Lead data scientist, AI solutions.*

*Outsourcing.* Case company III does not outsource the work. However, it uses the cloud-based AI platform from Microsoft for computational resources. Mostly internal developers use cloud services of Azure Machine learning platform, where they write their scripts and do necessary training of their models. In this case, developers from case company III are using a standard product of Microsoft without any specific collaboration agreement.

**Evaluation phase**

Most of the AI products at case company III become additional features to existing digital solutions offered to its clients. The developed AI product in the Evaluation phase is being integrated into a specific software. It follows further steps that apply to a software testing procedure, i.e., a set of practices that combines software development and IT operations – (DevOps). Further, an AI product needs to be brought up into the phases of CI (Continuous Integration) and CD (Continuous Development). An AI model will be automatically tested, validated, and deployed by software where it is integrated.

**Deployment**

Case company III follows the same principles and uses cloud solutions like the Azure Machine learning platform to host developed AI models. According to Lead data science, they deploy their models into AI cloud solutions and able to integrate them into a dedicated software, following its specific requirements.

### 5.4 Case Company IV

The empirical findings from the following case are based on interviews with the Head of the Research Group and the Lead Research Scientist from a multinational company that specializes within the energy sector in energy management, smart grid and power distribution equipment.

**Business Understanding**

The general approach of a case company IV is to consider AI technology as a tool to solve a problem and treat it the same way as any other technology. First, they start with the business problem at hand and then decide whether it is suitable or not to employ AI solutions. There are several main factors that are considered during the business problem investigation. Safety requirements (**human safety**) are always known at the beginning of the project, and there is a number of domains where safety plays a critical role, like a control system for a train, where the solution must work 100% of the time, or safety-critical devices like protection in the electrical power network. In these cases, applying AI solutions is problematic since, firstly, developers cannot guarantee performance on a fixed level. Secondly, obtaining a safety certification for AI-based technology is not yet possible. Another critical factor is **scale-up potential**. The approach in product development heavily depends on the future scale. For example, whether it would be a custom-made product for a particular customer which needs to be built only once or it would be millions of copies which would be sold on the market. If future product or solution can be sold in large quantities, then it is worth investing money in developing them in-house rather than going for partnership or off-the-shelf solutions since for case company IV it is always essential to keep the value-creating product entirely in-house.

> *"If you predict that clients will buy a million copies, then it is worth to put a million in your product development"- Head of Research Group*

Next critical factor is the **maturity of the technology**, like application of AI in a predictive maintenance or quality inspection. Those AI projects no longer require research but only building a product. Since there is a lot of accumulated knowledge and experience, the accurate cost estimation for this type of product development can be very straightforward. Confidence in the estimation of CTR (cost-time-resources) also affects how the company approaches the AI project execution for product development. **Timing** is another important criterion that influences the execution strategy for AI projects.

In case the product is for an **internal client**, the most important factor is how the developed product would improve the efficiency of the internal workflow.

In summary, the key criteria for the Business Understanding phase highlighted during interviews at case company IV are:

- *Is the application critical for human safety?*
- *What is the future scale-up?*

- *What is the maturity of the technology?*
- *Is timing critical?*

**Data Understanding and Preparation phase**

Case company IV, most of the time, performs the Data Understanding and Preparation phases in-house. The developers in case company IV first try to understand what data they need. Once they understand, they must understand how to get it. The in-house development team works with the client, whether internal or external, to see what data is available. For example, whether the data is in a stand-alone database or the cloud, whether there is a process in place for the data extraction, and whether authorization is required to access the data. Sometimes, clients have already extracted the data for developers upfront. In other cases, data still need to be gathered, and developers need to connect to the client's ecosystem by adding new sensors, for example, to get an insight on the process and collect necessary data points.

**Model Development phase**

*In-house approach.* For case company IV, in-house development is the approach when the product is part of the business value, especially if the future product is planned to be sold to customers in large quantities. Since most internal AI projects are based on mature technology and do not include research, it is realistic to do a very accurate estimation and allocate resources for the development. Case company IV has extensive experience and resources available for AI products development, and its general goal is to keep ownership of products and solutions that are part of the core business.

*Partnering with a third party.* If **timing is critical**, it might be easier to partner with a product developer first and then replicate the product in-house later, respecting third-party intellectual property rights. For AI projects where the technology readiness level of future products is low, the company's approach is to partner with academia or the government. If the researched technology is for industry benefit in general, they also might use a governmental funding.

> *"It is always better to have something in hand now than nothing just because customers will not wait for us to be ready. Then it is better to have at least part of the business than no business at all."*
> *- Head of Research Group*

*Outsourcing.* Case company IV has extensive knowledge and experience in AI project development and deployment. It also has a large pool of skilled personnel in various domains and geographic locations. Outsourcing part of the project to another company does not create for them any financial, quality or schedule benefits.

**Evaluation phase**

In case company IV, the research and development department focuses on the research in software and system processes, system engineering, software engineering and brings state-of-the-art prototypes to the development teams to work together for further integration of the AI product into digital solutions that company offers. Similar to case company III, AI products become part of the complex software, and a similar set of practices is applied for testing and their integration into an overall solutions.

**Deployment**

On the contrary to other researched companies, the case company IV has their own cloud solutions through which they deploy its AI products. Development of IoT (Internet of Things) platforms are part of the company's core business, and they always would use their own platforms to deploy and scale any AI products they create.

### 5.5 Case Company V

Case company V is a multinational company specializing in industrial software development and has two internal teams working in the field of AI. The first team is related to the research front end and consists of scientists and professionals with in-depth knowledge of industrial data science and domain expertise. The research team identifies the potential future product from various proposals and builds MVPs to hand it over to the production team. To ensure a lean process from ideation to productization, the research team continues to support the production team until product development is completed.

The empirical findings from this case company are based on an interview with a research team's Senior Vice President for Artificial Intelligence Technology.

**Business Understanding phase**

Case company V starts from the ideation phase, where the research team collects and brainstorms various ideas from academia partnerships, the internal product management team and customers. Further, the selected ideas are being researched for a short period to see whether they have the potential to become viable future products and to identify the risks for each idea implementation. To do so, the research team works closely with particular stakeholders and domain experts to get their feedback.

One of the first selection criteria is to address the **customer's pain points**. This criterion also helps not to defuse the focus by the trend in AI technologies but to stay focused on the customers' problem area.

According to the interviewer, building a separate product for each idea is not advisable. Too many products with different user interfaces might confuse the customer. Therefore, **usability** is, in the end, one of the key differentiators.

> *"It is very important (for customers) to have a holistic picture and simple usability because you can have the nicest feature, but if no one knows how to use it, people will use it in the wrong way."*
> *- Senior Vice President, AI Technology*

Another key factor is to have an **alignment with the company strategy**. Understanding what kind of value proposition a future product would enable and how it fits the business roadmap is essential. Therefore, it is important to bring research and development teams together at different levels and have a strategic alignment at the top level. Ultimately, it is always a trade-off between risk and reward. Risk, in this case, is related to the time and resources invested in the proof of concept and development.

After filtering the initial ideas based on the key factors and building proof of concepts, the research team hands it over to production. To ensure that the gap between research and actual implementation is adequately addressed and to mitigate the risks, the research team keep working together with the operational team and customers, verifying initial assumptions via a constant iteration loop till the deployment of the product.

It is worth mentioning that the research team is always working on several ideas in parallel to de-risk the invested time and resources. That can be considered as benefit of large companies vs start-ups who must be very focused on one idea without having the luxury to invest simultaneously in various different projects.

In summary, the key criteria for the Business Understanding phase highlighted during interviews at case company V are:

- *Which market is targeted?*
- *What type of customers are there?*
- *Is the goal clear?*
- *What are the investments?*

**Data Understanding and Preparation phase**

To address customers' actual problems, the research team stays connected to the company's customers. A lot depends on the relationship with customers and the way they are engaged in the research. In the case company V, it happens through customer workshops, innovation workshops, brainstorming, and through these peer-to-peer relations.

It is important to establish direct contact with customers and constantly get their feedback. However, there might not always be possible to have brainstorming activities together due to various reasons. In some market areas, the customers can also be potential competitors. Therefore, early engagement and trusted relationships are essential. The data can be exchanged in both directions. However, in some cases, it is challenging to have direct access to the data due to security and privacy reasons, for example, in the case of smart grid companies. It is essential to handle data appropriately and keep customers aware of what it is used for; to ensure that the data access is adequately controlled and deleted after research is completed.

Case company V also has engineering as a core part of its business that focuses on automating manual work, for example, like labelling, as much as possible. Therefore, all data preparation is also done internally.

**Model Development phase**

<u>*In-house approach.*</u> In general, case company V might consider different approaches. However, given that it is a software company, it is essential to develop the core differentiator project internally. Therefore, case company V invests significantly in people resources.

> *"You can use acceleration in areas that are not your differentiator and where you don't want to create a market advantage. Using available services and products can give you scale and speed, but if you are trying to outsource your main market differentiator, you have nothing."- Senior Vice President for AI Technology*

<u>*Partnering with a third party*</u>. The case company V would partner with external companies in the use cases when it is not related to the core competency. The company is eager to keep **IP** on its market differentiators and collaborate with others only on projects outside of its core domain.

Case company V as well collaborates extensively with **academia** in different ways. For example, through various academic partnership programs, internships and regular meetings with universities. Although, they do not involve academia in high-IP product development.

<u>*Outsourcing.*</u> Case company V would only outsource work which is outside its core domain to accelerate time to step into a market and to scale. For example, it has a strong relationship with big cloud providers since it helps to speed up the internal innovation process.

**Evaluation phase**

Case company V has a formal process to ensure the desired quality and to confirm that the product meets its specification. In addition, the company selects a number of important clients in trial product testing to get their feedback.

**Deployment**

Similar to case company IV, case company V prefers to use its own environment to develop and deploy its AI products because it is considered as a financially attractive way to scale up the production in a longer term.

## 6. RESEARCH SUMMARY

To present findings from all case companies, we structure them as a decision tree on **Fig. 9.** The key factors revealed during case studies are the decision tree nodes which are highlighted in **bold** in section 5. Contrary to **Fig. 8**, which summarizes key factors in developing and deploying AI solutions at an EPC contractor (case company I) which is in the initial phase of introducing AI into its business, the final framework in **Fig. 9** reflects the findings from the case companies that are already benefitting from integrating AI in their operations and final products for their clients.

As shown in **Fig. 9**, the first step is to distinguish the AI project purpose, i.e., whether it is aligned with a business strategy. If the project is not aligning with the company roadmap, it is likely to be filtered out in the initial ideation phase. Next, it is important to distinguish whether an AI product is for an **internal** or **external client**.

The following important factor is whether the future product is the potential **business differentiator**. Only in this case, the AI project for an external client passes through the gate review in the Business Development phase. If the product is **highly specialized** and **critical to the market**, most probably, collaboration with a specialized company is the best approach. However, in this case, special attention to IP rights is required. If **IP** rights are prioritized, it is advised to keep the development in-house.

The next important factor is whether the future product has **scale-up potential**. If so, the product serves the business long-term and is worth a larger investment to be developed in-house completely.

**The general maturity of the industry** is considered too, since re-investing in the product which already exists is not the intention, and it makes sense to re-consider the business model and look for the potential benefits using existing products or collaboration with companies who developed similar product earlier.

**The high technology readiness** level assumes that the product exists on the market but not necessarily is utilized widely by the industry. In this case, collaboration with the product developer could be an optimal approach.

In case of **low technology readiness level** but the product is considered as a potential business differentiator on the market and has a scale-up opportunity, it is worth investing in internal research and collaborating with academia.

**For an internal client**, the above factors are not crucial. Usually, internal clients are not a core business, i.e., HR or legal department. Therefore, it is recommended to look for existing off-the-shelf solutions or outsource the work where it is possible. In case of the technology readiness level is low or the workflow the future product is supposed to support is very specialized within the company, then internal research might be the best approach.

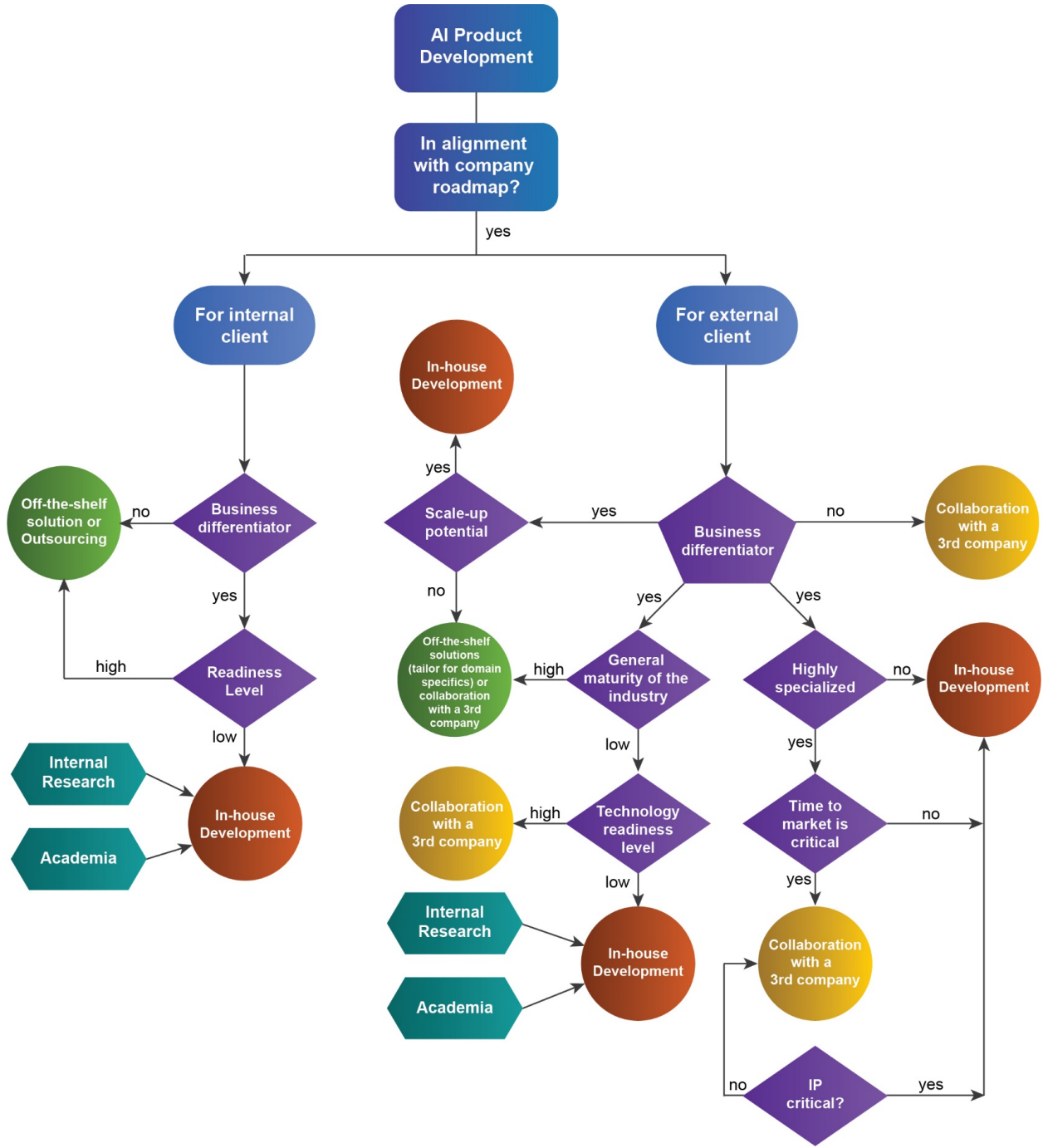

**Fig. 9: Overall Framework for the most optimum approach selection for AI project development and deployment**

## 7. DISCUSSION

In this research, we studied four approaches for AI technologies' industrial development and deployment: (1) in-house development, (2) collaboration with a third-party specialized in AI, (3) utilization of off-the-shelf solutions and (4) outsourcing the work to a third party.

The empirical results derived from the development and deployment of AI projects at Company I, as detailed in subsection 5.1.1, have been analyzed to address Research Questions 1.1 and 1.2. From these results, we constructed a framework that identifies the critical factors in developing and deploying AI solutions within an EPC company with limited experience in this field (Fig. 8).

To validate these findings and enrich the framework, we extended our research to include interviews with solutions. The empirical data gathered from these interactions, detailed in subsections 5.2 - 5.5, laid the groundwork for developing an overall framework, which also addresses Research Questions 2.1 and 2.2. This framework is presented in Section 6, "Research Summary" (Fig. 9). It shows the evolution of key factors as companies enhance their proficiency in integrating AI solutions into their business operations.

As mentioned in Section 3, the research on industrial ML/DL application in engineering is limited (Borges et al., 2021), (Kitsios et al., 2021). In contrast, a lot of research can be found on IT applications. Similar strategies, such as outsourcing or collaborating with specialized third parties, have been explored for crucial IT applications, emphasizing major opportunities and risks (Verner and Abdullah, 2012), (Nie and Hammouda, 2017). The novelty of our research is that we enriched previous studies by focusing on AI applications. We examined how businesses in the EPC industry within the energy sector integrate AI with their strategies. This included studying a case company just starting with AI solutions and other firms in the same industry that have already established a mature AI presence in their operations. The research outcome allows us to see how the criteria for approach selection for AI solutions development and deployment evolve as companies mature in their AI integrations. It also helps to bridge the gap in academic research by sharing practitioners' insights on AI project management and strategies for integrating AI technology into businesses.

To enrich the discussion, we compared our findings with academic studies on factors influencing technology adoption, development, deployment, and integration:

**Common Factors**

*Business Strategy Alignment:* Our framework (Fig. 9) and Kitsios et al., 2021's findings underscore the significance of aligning AI with company goals, mirroring Borges et al., 2021's perspective.

*Innovation Potential:* We emphasize scalability and unique business offerings, while Kitsios et al., 2021 focus on innovative potential and human-machine collaboration.

*Readiness and Maturity:* Our framework (Fig. 9) points to industry and technology maturity. Concurrently, Ridwandono and Subriadi, 2019 discuss organizational flexibility, and Kitsios et al., 2021 talk about the strategic utilization of AI in terms of competitive and cognitive strategies.

*Collaboration:* Both the research and works by Kitsios et al., 2021 and Bertolini et al., 2021 emphasize collaboration with third parties, while our overall framework (Fig. 9) further details such collaboration with a third parties in various contexts.

**Different Factors**

*Human Interaction with AI:* Kitsios et al., 2021 discuss human feelings, attitudes, and motivations to interact with cognitive technologies.

*Concerns about AI:* The paper by Bertolini et al., 2021 mentions concerns about job loss and unemployment due to AI.

*Challenges in Deployment:* The paper by Paleyes et al., 2021 discusses the challenges in deploying ML in industries and suggests a systematic literature review to address these.

*Resources for AI Development:* Mikalef et al., 2019 focus on the importance of various resources, such as technical skills, managerial skills, and a data-driven culture. Our findings don't specify these resources but rather focus on decision points.

*Fear and Understanding of AI:* Borges et al., 2021 highlight the fear of job elimination and the need for understanding and explaining AI behavior. These are not directly addressed in our work.

In summary, while our research offers a holistic decision-making roadmap for AI product development and deployment, it might not reflect every academic highlight. Nevertheless, the potential implications of our work include providing valuable insights into decision-making processes at the leadership level within organizations regarding AI integration, which was not found in related work from academia. By focusing on business leaders - those at the helm of decision-making and influential in shaping corporate business strategies regarding AI integration - this study has provided a unique viewpoint into these processes. The findings also contribute to a more comprehensive understanding of AI adoption and implementation across different industries. In our opinion, future work could benefit from incorporating insights on human interaction with AI, challenges in deployment, and specific resources necessary for successful AI development and deployment. We also propose a future study that involves post-evaluation or validation of the overall framework derived from this research. This follow-up study could involve a comprehensive survey of a larger group of professionals in the field, spanning diverse organizations and roles.

## 8. THREATS TO VALIDITY

Several factors may threaten the internal validity of this case study-based research. First, the selection of case companies and interviewees could potentially introduce bias. Second, the interpretation of data, if subjected to researcher bias, may skew the findings. Additionally, the methodological limitations inherent to case study research require careful consideration. While it provides valuable insights into human perspectives on the use and development of computer-based information systems, this method is time-consuming and expensive, which could potentially limit the scope of the research (Walsham, 1995). The uniqueness of each case can complicate the generalization of the findings. Finally, the potential influence of researchers' biases may affect the research outcome, requiring cautious interpretation of the findings (Dubé and Paré, 2003).

With respect to external validity, taking an entire company as one single case might threaten the validity of the findings. However, the interviewers represent decision-makers for their companies within the AI field particularly. Therefore, the data gathered for each case company provide unique insights that are general for developing and deploying particularly AI projects within each case company. The researchers also acknowledge the limitations posed by inaccessible data from other industry companies but assert that their findings can still offer valuable perspectives on the industry as a whole.

Other threats to the validity of this research include the potential for the case companies to be unrepresentative of the population of companies in the industry. Future research must acknowledge these potential threats and take necessary measures to mitigate their effects to ensure accurate and reliable conclusions.

## 9. CONCLUSION

In this paper, we analyzed in detail the insights from practitioners - one of the biggest EPC contractor worldwide and large EPC vendors that are experienced in project execution for developing and deploying AI-based products.

The analysis based on AI project development and deployment experience in the case company I represents the most optimum approach of the large corporation that started integrating AI into its business and has not yet accumulated sufficient experience. The analysis of the empirical data received through interviewing the EPC vendors who are working in the same industry shows how these key factors are transformed as the business becomes more mature in integrating AI into its workflow, products, and solutions they offer. Those interviews provide insight into how companies approach projects for AI product development and what factors can affect their choices. All interviewed people are the decision-makers in choosing the strategy of integrating AI into their organization. The overall framework that we developed based on this analysis covers the entire life cycle of building AI solutions, starting from initial business understanding until deployment and further evolution. The fact that it might be similar to many other businesses and technologies as well demystifies AI, especially for EPC companies that are beginning to consider the integration of AI into their operations. These insights can be shared with businesses and academia, adding value to both sectors and helping to facilitate AI integration into EPC business within the energy sector.

Our findings, while specifically focusing on the EPC industry, bear relevance to AI product development in other business domains as well. The process of AI integration, the factors influencing the choice of different AI strategies, and the impact on long-term business strategy are areas that cut across various industries. The insights we have gained about the decision-making processes and AI implementation challenges can provide valuable guidance to other sectors beyond EPC.

It would be advisable as future work to compare our findings with those from studies on AI product development in other sectors to identify shared challenges, strategies, and trends. This, in turn, can contribute to a more noted that while the overarching themes could be similar, the specific approaches and solutions may differ due to the unique characteristics, regulations, and dynamics of each industry. Therefore, while our findings can offer general guidance, they may need to be adapted or nuanced to fit the specific circumstances of other business domains.

## ACKNOWLEDGEMENTS

This research has been supported by McDermott International Inc and Software Center (Gothenburg, Sweden).

# References

Abioye, S.O., Oyedele, L.O., Akanbi, L., et al. Artificial intelligence in the construction industry: A review of present status, opportunities and future challenges. Journal of Building Engineering. 2021;44, https://doi.org/10.1016/j.jobe.2021.103299

Ahmad, T., Zhang, D., Huang, C., Zhang, H., Dai, N., Song, Y., Chen, H.; Artificial intelligence in sustainable energy industry: Status Quo, challenges and opportunities, Journal of Cleaner Production, Volume 289, 2021, 125834, ISSN 0959-6526, https://doi.org/10.1016/j.jclepro.2021.125834.

Ahmad, T., Zhu ,H., Zhang, D., Rasikh, T., Bassam, A., Ullah, F., Alghamdi, A., Alshamrani, S. (2022). Energetics Systems and artificial intelligence: Applications of industry 4.0. Energy Reports. 8. 334-361. https://doi.org/10.1016/j.egyr.2021.11.256

Alves, T., Ravaghi, K., Needy, K. (2016). Supplier Selection in EPC Projects: An Overview of the Process and Its Main Activities. 209-218. http://dx.doi.org/10.1061/9780784479827.022

Amershi, S., Begel, A, Bird, C., DeLine, R., Gall, H., Kamar, E., Nagappan, N. Nushi, B., Zimmermann, T., 2019. Software Engineering for Machine Learning: A Case Study. IEEE/ACM 41st International Conference on Software Engineering: Software Engineering in Practice (ICSE-SEIP), pp. 291-300, https://doi.org/10.1109/ICSE-SEIP.2019.00042.

Arpteg, A., Brinne, B., Crnkovic-Friis, L., Bosch, J. (2018). Software engineering challenges of deep learning. In 2018 44th Euromicro Conference on Software Engineering and Advanced Applications (SEAA) (pp. 50-59). IEEE, http://dx.doi.org/10.1109/SEAA.2018.00018.

Ashmore, R., Calinescu, R., & Paterson, C., 2019. Assuring the Machine Learning Lifecycle: Desiderata, Methods, and Challenges. https://doi.org/10.48550/arXiv.1905.04223.

Balakrishnan, T., Chui, M., Hall, B., Henke, N., November 2020. McKinsey Digital, McKinsey Analytics, Global survey: The state of AI in 2020, Global survey: The state of AI in 2020 | McKinsey

Balmer, R.E. Levin, S.L., Schmidt, S., 2020. Artificial Intelligence Applications in Telecommunications and other network industries, Telecommunications Policy, Volume 44, Issue 6, 2020, 101977, ISSN 0308-5961, https://doi.org/10.1016/j.telpol.2020.101977.

Baron, H., 2018. The Oil & Gas Engineering Guide, third edition . Editions Technip, 300 p.

Bertolini, M., Mezzogori, D., Neroni, M., Zammori, F., 2021. Machine Learning for industrial applications: A comprehensive literature review, Expert Systems with Applications, Volume 175, 2021, 114820, ISSN 0957-4174, https://doi.org/10.1016/j.eswa.2021.114820.

Blackridge Research & Consulting, 2022 What is an EPC Contract? Here's what you need to know. https://www.blackridgeresearch.com/blog/what-is-an-epc-contract. (accessed 23 February 2023).

Blanco, J.L., Funch, S., Parsons, M., Ribeirinho, M.J. Artificial intelligence: Construction technology's next frontier. 2018. https://www.mckinsey.com/capabilities/operations/our-insights/artificial-intelligence-construction-technologys-next-frontier

Borges, A.F.S., Laurindo, F.J.B., Spínola, M.M., Gonçalves, R.F., Mattos, C.A.; 2021. The strategic use of artificial intelligence in the digital era: Systematic literature review and future research directions, International Journal of Information Management, Volume 57, 2021, 102225, ISSN 0268-4012, https://doi.org/10.1016/j.ijinfomgt.2020.102225.

Bose, B.K, 2017 Artificial Intelligence Techniques in Smart Grid and Renewable Energy Systems—Some Example Applications, in Proceedings of the IEEE, vol. 105, no. 11, pp. 2262-2273, https://doi.org/10.1109/JPROC.2017.2756596

Braun, V., Clarke, V., 2006 Using thematic analysis in psychology, Qualitative Research in Psychology, 3:2, 77-101, https://doi.org/10.1191/1478088706qp063oa

Chui, M., Hall, B., Singla, A., Sukharevsky, A. 2021. McKinsey Digital, McKinsey Analytics, Global survey: The state of AI in 2021, Global survey: The state of AI in 2021 | McKinsey

Collins, C. , Dennehy, D., Conboy, K., Mikalef, P., 2021. Artificial intelligence in information systems research: A systematic literature review and research agenda, International Journal of Information Management, Volume 60, 2021, 102383, ISSN 0268-4012, https://doi.org/10.1016/j.ijinfomgt.2021.102383.

Dimensional research (supported by ALEGION), 2019. Artificial Intelligence and Machine Learning Projects Are Obstructed by Data Issues Global Survey of Data Scientists, AI Experts and Stakeholders, Dimensional Research Machine Learning Report, May 2019

Dubé, L., Paré, G., 2003. Rigor in Information System Positivist Case Research. MIS Quarterly, 27(4), 597–635. http://dx.doi.org/10.2307/30036550

Dzhusupova, R., Bosch, J., Holmström Olsson, H, 2022. The Goldilocks Framework: Towards Selecting the Optimal Approach to Conducting AI Projects, 2022 IEEE/ACM 1st International Conference on AI Engineering – Software Engineering for AI (CAIN), Pittsburgh, PA, USA, pp. 124-135, https://doi.org/10.1145/3522664.3528595.

Dzhusupova, R., Bosch, J., Holmström Olsson, H., 2022. Challenges in developing and deploying AI in the engineering, procurement and construction industry. 2022 IEEE 46th Annual Computers, Software, and Applications Conference (COMPSAC). https://doi.org/10.1109/COMPSAC54236.2022

Easterbrook, S., Singer, J., Storey, M-A., Damian, D., 2007. Selecting Empirical Methods for Software Engineering Research. Guide to Advanced Empirical Software Engineering, pp. 285-311. http://dx.doi.org/10.1007/978-1-84800-044-5_11

Enholm, I.M. & Papagiannidis, E., Mikalef, P., Krogstie, J., 2021. Artificial Intelligence and Business Value: a Literature Review. Information Systems Frontiers 24, pp1709–1734. http://dx.doi.org/10.1007/s10796-021-10186-w.

ESFC Company, EPC Contractor: financing, engineering and construction. https://esfccompany.com/en/articles/engineering/epc-contractor-engineering-and-construction/. Accessed 23 February 2023.

Fahle, S., Prinz, C., Kuhlenkötter, B., 2020. Systematic review on machine learning (ML) methods for manufacturing processes – Identifying artificial intelligence (AI) methods for field application, Procedia CIRP, Volume 93, 2020, Pages 413-418, ISSN 2212-8271, https://doi.org/10.1016/j.procir.2020.04.109

Fayyad, U., Piatetsky-Shapiro, G., Smyth, P, 1996. The KDD Process for Extracting Useful Knowledge from Volumes of Data; COMMUNICATIONS OF

Forootan, M.M., Larki, I., Zahedi, R., Ahmadi, A.. 2022. Machine Learning and Deep Learning in Energy Systems: A Review Sustainability 14, no. 8: 4832. https://doi.org/10.3390/su14084832

Gartner, Stamford, Conn., January 21, 2019. Press Release STAMFORD, Conn., Gartner Survey Shows 37 Percent of Organizations Have Implemented AI in Some Form, https://www.gartner.com/en/newsroom/press-releases/2019-01-21-gartner-survey-shows-37-percent-of-organizations-have.

Ghoddusi, H., Creamer, G., Rafizadeh, N., 2019. Machine Learning in Energy Economics and Finance: A Review. Energy Economics. 81. http://dx.doi.org/10.1016/j.eneco.2019.05.006.

Hamm, P., Klesel, M., 2021 Success Factors for the Adoption of Artificial Intelligence in Organizations: A Literature Review . AMCIS 2021 Proceedings. 10. https://aisel.aisnet.org/amcis2021

Hanga, K.M., Kovalchuk, Y., 2019. Machine learning and multi-agent systems in oil and gas industry applications: A survey. Comput. Sci. Rev., 34. https://doi.org/10.1016/j.cosrev.2019.08.002

Hansen, S., 2015. Study on the Management of EPC Projects. International Journal of Civil, Structural, Environmental and Infrastructure Engineering Research and Development (IJCSEIERD). 2015;5(3):11-22. https://www.researchgate.net/publication/337240567_Study_on_the_Management_of_EPC_Projects

Hill, C., Bellamy, R., Erickson, T., Burnett, M., 2016. Trials and tribulations of developers of intelligent systems: A field study, 2016 IEEE Symposium on Visual Languages and Human-Centric Computing (VL/HCC), pp. 162-170, https://doi.org/10.1109/VLHCC.2016.7739680

IEC 61508-1. 2010, International Standard. Functional safety of electrical/electronic/programmable electronic safety-related systems - Part 1: General requirements. https://webstore.iec.ch/publication/5515

Irena, World Energy Transition Outlook 2022, World Energy Transitions Outlook 2022 (irena.org) (accessed 24 February 2023)

Khan, Z., Hussain, T., Baik, S., 2022. Boosting energy harvesting via deep learning-based renewable power generation prediction. Journal of King Saud University – Science, vol. 34, Issue 3. https://doi.org/10.1016/j.jksus.2021.101815

Kiger, M., Varpio, L., 2020. Thematic analysis of qualitative data: AMEE Guide No. 131, Medical Teacher, 42:8, 846-854, https://doi.org/10.1080/0142159x.2020.1755030

Kitsios, F., Kamariotou, M., 2021. Artificial Intelligence and Business Strategy Towards Digital Transformation: A Research Agenda. Sustainability. 13 (4) 2025. http://dx.doi.org/10.3390/su13042025.

Koroteev, D., Tekic, Z., 2021. Artificial intelligence in oil and gas upstream: Trends, challenges, and scenarios for the future, Energy and AI, Volume 3, 2021, 100041, ISSN 2666-5468, https://doi.org/10.1016/j.egyai.2020.100041

Lee, J., Kim, Y. Lee, S-Y., Lee, E-B., 2020. Development of AI-Based Engineering Big Data Integrated Analysis System for Decision-Making Support in the Engineering-Procurement-Construction(EPC) Industry. Proceedings of the 37th International Conference of CIB W78, Sao Paulo - online, BR, 18-20 August, pp. 58-68 (ISSN: 2706-6568), http://itc.scix.net/paper/w78-2020-paper-005

Mikalef, P., Boura, M., Lekakos, G., Krogstie, J., 2019. Big data analytics and firm performance: Findings from a mixed-method approach. Journal of Business Research. 98. 261-276. https://doi.org/10.1016/j.jbusres.2019.01.044

MSCI, 2020 GLOBAL INDUSTRY CLASSIFICATION STANDARD (GICS®) METHODOLOGY, Guiding Principles and Methodoloty for GICS. https://www.msci.com/documents/1296102/11185224/GICS+Methodology+2020.pdf

Nie, E., & Hammouda, I., 2017. An Exploratory Study on Strategic Software Development Outsourcing. 2017 IEEE 12th International Conference on Global Software Engineering (ICGSE), pp. 106-115, https://doi.org/10.1109/ICGSE.2017.10.

Paleyes, A.. Urma, R-G., Lawrence, N., 2020. Challenges in Deploying Machine Learning: a Survey of Case Studies, Cornell University, The ML-Retrospectives, Surveys & Meta-Analyses Workshop, NeurIPS 2020; v3, 19 May 2022, https://doi.org/10.48550/arXiv.2011.09926.

Panikkar, R., Saleh, T., Szybowski, M., Whiteman, R. September 2021, article. Operationalizing machine learning in processes, McKinsey&Company. https://www.mckinsey.com/capabilities/operations/our-insights/operationalizing-machine-learning-in-processes

Pappas, I.O.; Mikalef, P.; Giannakos, M.N.; Krogstie, J.; Lekakos, G., 2018. Big data and business analytics ecosystems: Paving the way towards digital transformation and sustainable societies. Information Systems and e-Business Management 16, 479–491, https://doi.org/10.1007/s10257-018-0377-z

Park, M-J., Lee, E-B., Lee, S-Y., Kim, J-H., 2021. A Digitalized Design Risk Analysis Tool with Machine-Learning Algorithm for EPC Contractor's Technical Specifications Assessment on Bidding. Energies. 14(18):5901, http://dx.doi.org/10.3390/en14185901.

Putra, G., Triyono, R., 2015. Neural Network Method for Instrumentation and Control Cost Estimation of the EPC Companies Bidding Proposal. Procedia Manufacturing, 4. 98-106. https://doi.org/10.1016/j.promfg.2015.11.019

Raji, I.D., Smart, A., White, R., Mitchell, M., Gebru, T., Hutchinson, B., Smith-Loud, J., Theron, D., Barnes, P., 2020. Closing the AI accountability gap: defining an end-to-end framework for internal algorithmic auditing. Proceedings of the 2020 Conference on Fairness, Accountability, and Transparency, https://doi.org/10.48550/arXiv.2001.00973.

Ridwandono, D., Subriadi, A., 2019. IT and Organizational Agility: A Critical Literature Review. Procedia Computer Science. 161. 151-159. https://doi.org/10.1016/j.procs.2019.11.110

SAIs, Supreme Audit Institutions of Finland, Germany, the Netherlands, Norway and the UK, November 2020. Auditing machine learning algorithms. A white paper for public auditors. https://www.auditingalgorithms.net/index.html

Saunders, M., Lewis, P., Thornhill, A., 2015. Research methods for business students, seventh edition. Pearson,

Sculley, D., Holt, G., Golovin, D., Davydov, E., Phillips, T., Ebner, D., Chaudhary, V., Young, M., Dennison, D., 2015. Hidden Technical Debt in Machine Learning Systems. NIPS. 2494-2502, http://papers.nips.cc/paper/5656-hidden-technical-debt-in-machine-learning-systems.pdf.

Sholeh, M.N., Fauziyah S., 2018. Current state mapping of the supply chain in engineering procurement construction (EPC) project: a case study. MATEC Web of Conferences.195:06015, http://dx.doi.org/10.1051/matecconf/201819506015.

Soiferman, L.K., 2010. Compare and Contrast Inductive and Deductive Research Approaches. ERIC Number: ED542066; 2010.

Sourkouhi, Z., 2013. Strategic management; concepts, benefits and process. IOSR Journal of Business and Management. 13. 61-64. http://dx.doi.org/10.9790/487X-1336164.

Statista, https://www.statista.com/statistics/217556/percentage-of-gdp-from-energy-in-selected-countries/ (accessed 23 February 2023)

Statista, Energy investments globally by sector | Statista (accessed 23 February 2023)

Studer, S., Bui, B., Drescher, C., Hanuschkin, A., Winkler, L., Peters, Steven & Müller, Klaus-Robert. (2021). Towards CRISP-ML(Q): A Machine Learning Process Model with Quality Assurance Methodology. Machine Learning and Knowledge Extraction. 3. 392-413. https://doi.org/10.3390/make3020020.

Tab M., Buck, A., Sharkey, K., December 2021. What is the Team Data Science Process? What is the Team Data Science Process? - Azure Architecture Center | Microsoft Learn (accessed 23 February 2022).

Venkataraman, V., Narang, A., Vakil, T., 2022 PricewaterhouseCoopers Private Limited https://www.pwc.in/assets/pdfs/emerging-tech/5g-transforming-the-engineering-procurement-and-construction-epc-industry.pdf

Verner, J.M., Abdullah, L.M., 2012. Exploratory case study research: Outsourced project failure. Information and Software Technology. 54, 8, pp. 866–886. https://doi.org/10.1016/j.infsof.2011.11.001

Walsham, G. Interpretive case studies in IS research: nature and method. Eur J Inf Syst 4, 74–81 (1995). https://doi.org/10.1057/ejis.1995.9

Widodo, D., Iksan, N., Udayanti, E., Djuniadi, 2021. Renewable energy power generation forecasting using deep learning method. IOP Conference Series: Earth and Environmental Science. 700. 012026, http://dx.doi.org/10.1088/1755-1315/700/1/012026.

Wiggers, K., 2019. IDC: For 1 in 4 companies, half of all AI projects fail, VentureBeat, July 8, 2019, based on IDC's Artificial Intelligence Global Adoption Trends & Strategies report. IDC: For 1 in 4 companies, half of all AI projects fail | VentureBeat

Wirth, R. & Hipp, J., 2000. CRISP-DM: Towards a standard process model for data mining. Proceedings of the 4th International Conference on the Practical Applications of Knowledge Discovery and Data Mining http://www.cs.unibo.it/~montesi/CBD/Beatriz/10.1.1.198.5133.pdf.